\documentclass[useAMS,usenatbib]{mn2e} 
\usepackage[varg]{txfonts}
\usepackage{natbib}
\usepackage{graphicx}
\usepackage{txfonts}

\title[Generation of inclined protoplanetary discs through stellar flybys]{Generation of highly inclined protoplanetary discs through single stellar flybys}
\author[M. Xiang-Gruess]{M. Xiang-Gruess$^{1}$ \thanks{E-mail:
mxianggruess@mpifr-bonn.mpg.de }  \\
$^{1}$ Max-Planck Institut f\"ur Radioastronomie, Auf dem H\"ugel 69, D-53121 Bonn, Germany\\
}

\begin{document}

\date{Accepted . Received ; }

\pagerange{\pageref{firstpage}--\pageref{lastpage}} \pubyear{2015}
\maketitle

\label{firstpage}

\begin{abstract}

We study the three-dimensional evolution of a viscous protoplanetary disc which is perturbed by a passing star on a parabolic orbit. The aim is to test whether a single stellar flyby is capable to excite significant disc inclinations which would favour the formation of so-called misaligned planets. 
We use smoothed 
particle hydrodynamics to study inclination, disc mass and angular momentum changes of the disc for passing stars with different masses. We explore different orbital configurations for the perturber's orbit to find the parameter spaces which allow significant disc inclination generation.
Prograde inclined parabolic orbits are most destructive leading to significant 
disc mass and angular momentum loss. In the remaining disc, the final disc inclination is only below 20\degr. This is due to the removal of disc particles which have experienced the strongest perturbing effects.
Retrograde inclined parabolic orbits are less destructive and can generate disc inclinations up to $60\degr$. 
The 
final disc orientation is determined by the precession of the disc angular momentum vector about the perturber's orbital angular momentum vector and by disc orbital inclination changes.

We propose a sequence of stellar flybys for the generation of misalignment angles above $60\degr$.  
The results taken together show that stellar flybys 
are promising and realistic for the explanation of misaligned Hot Jupiters with misalignment angles up to 60\degr. 

\end{abstract}


\section{Introduction}

In the past few decades, the number of detected extrasolar planets around other main sequence stars has increased dramatically, so that  at the time of writing,  more than 1000 confirmed planets have been discovered among which $\sim 10\%$ have an observed measure of the angle between their orbital plane and the equatorial plane of their host star \citep[e.g.][]{Win2007, Tri2010}. Well studied  planet formation  scenarios  such as  core accretion  \citep{Miz1980, Pol1996} or disc fragmentation \citep{May2002} involve the planet forming in a disc with the natural expectation that the orbit should be coplanar. However, Rossiter McLaughlin measurements  of close orbiting Hot Jupiters indicate that around $40\%$ have angular momentum vectors significantly misaligned with the angular velocity vector of the central star \citep[e.g.][]{Tri2010, Alb2012}. 
Recently, new techniques have also been applied on the detection of the misalignment angles such as a technique involving asteroseismology \citep[e.g.][]{Hub2013} or star-spot measurements \citep[][]{Hir2012}.
Although firstly discovered in systems involving single Hot Jupiters, new observations indicate that significant misalignments can also occur in multi-planetary systems \citep[e.g.][]{Hub2013}.

 As the stellar and disc angular velocities are naturally expected to be aligned, the observational results would imply an inclination of the planetary orbits relative to the nascent protoplanetary disc. 

This has been underlined by several recent studies of the interactions of planets with a range of initial orbital inclinations with respect to a  disc \citep[e.g.][]{Cre2007, Mar2009, Bit2011}. In these works the evolution of planets with masses  up to a maximum of  $1 M_J$ with different initial eccentricities and relatively small  initial inclinations up to a maximum of $15\degr$ interacting with three-dimensional isothermal and radiative discs are considered.  They showed damping for both eccentricity and inclination with the planets circularising  in the disc after a few hundred orbits.
Planets with higher inclinations have also been studied by e.g. \citet{Rei2012}  and  \citet{Tey2013}.
A detailed survey of a range of planet masses and the full range of orbital inclinations interacting with a disc  has been presented by \citet{Xia2013}. The timescale of realignment with the gas disc was found to be  comparable with the disc  lifetime for very high inclinations $ >70^{\circ}$ and planet masses of  one Jupiter mass.  For smaller initial relative inclinations  and/or  larger planet masses,  the timescale for  realignment is shorter than the disc lifetime. These results taken together imply that there is  a need for an explanation for the origin  of  misaligned giant planets, if it is assumed that the discs angular momentum vector is always  aligned with the spin axis of the central star.

Several  scenarios have been proposed to explain the origin of misalignments between the planetary orbital angular  momentum vector  and the stellar rotation  axis  as well as produce close in giant planets. The first involves excitation  of very high eccentricities, either through  the Lidov-Kozai
effect induced by the  interaction with a distant companion \citep[e.g.][]{Fab2007, Wu2007}, or through planet-planet scattering or chaotic interactions \citep[e.g.][]{Weid1996,Rasio1996,Pap2001,Nag2008}. When the planet attains a small enough pericentre distance, this is then followed by orbital circularization due to tidal interaction with the central star, leaving the planet on a close inclined  circular orbit.

A method for producing high orbital inclinations with respect to the stellar equator has also been indicated by \cite{Tho2003}, who  have studied the evolution of two giant planets in resonance, adopting an approximate analytical expression for the influence of a gas disc in producing  orbital migration. Their calculations suggest that resonant inclination excitation can occur when  the eccentricity of the inner planet reaches a threshold value  $\sim  0.6$. Inclinations gained by the resonant pair of planets can reach values up to $\sim 60\degr$. 

However in  a recent study,   \citet{Daw2013} find that large eccentricities,  that are expected to  be associated with such interactions for  orbits too wide to be affected by stellar tides,   are seen  mostly
in metal rich systems. Accordingly they conclude  that  gentle  disc migration and planet-planet scattering must  both operate during the early evolution of giant planets. Only systems packed with giant planets, which most easily form around metal rich stars, can produce large eccentricities through scattering.
They also note a lack of a correlation between spin-orbit misalignment and metallicity that  may indicate the operation of such a mechanism. 


An explanation for the origin of  misaligned planets, that is consistent  with formation in  and inward migration driven by a disc may be   connected with the possibility that the orientation of the disc changes, possibly due  either to  a lack of orbital  alignment of the source material or to gravitational encounters with passing stars. This may occur either before or after  planet formation  \citep[e.g.][]{Bat2010,Thi2011}.
\citet{Thi2011}  showed that the capture of gas from accretion envelopes by  a protoplanetary disc could cause it to become significantly misaligned with respect to its original plane. It is of   interest to remark that  misaligned gas discs can easily be produced  when  interactions  with the environment out of which they form are considered. However, this result has been questioned recently by \citet{Pic2014} who studied the 3D evolution of circumstellar discs during stellar flybys including mass transfer between the two gas discs around the stars. In their simulations, disc inclinations of $<10\degr$ were only generated. 

  The effects of single close encounters on circumstellar discs can be very strong as the perturbing star can destroy the global structure and dynamics of a disc. These gravitational effects of a close stellar encounter on a circumstellar disc have been studied analytically as well as numerically by many groups in the past years \citep[e.g.][]{Cla1993, Kor1995, Kob2001, Pfa2003, Fra2009}. For different perturber orbits, these studies have mainly concentrated on equal-mass stars, disc mass loss and disc sizes. 
 
\citet{Kob2001} studied the eccentricity and inclination changes of test particles for inclined perturber orbits through orbital integration and analytical calculations. Though, the viscosity as well as self-gravity of the disc have not been taken into account making the results only partly applicable to realistic protoplanetary discs. 
One major lack of above studies is the assumption of equal-mass stars. 
In this regard we note that in general, protostars are not  formed in isolation, but are  bound in stellar clusters where gravitational interactions  are common. In stellar clusters, there is a global function describing the distribution of the stellar masses - the initial mass function (IMF) \citep[e.g.][]{Kro2002, Cha2003}. Taking into account the mass distribution in a stellar cluster, it can be very common that close encounters occur for different mass stars. 
So far, many studies of protoplanetary discs and planetary systems affected by single stellar flybys have investigated a parameter range spanning the entire IMF \citep[e.g.][]{Mal2011, Bre2014}. 

Several groups have performed N-body simulations of stellar clusters \citep[e.g.][]{Pfa2008, Spu2009, Mal2007, Mal2011, Olc2012, Cra2013, Vin2015} in order to study rates and properties of close encounters and their influences on disc sizes and mass loss, but also on planetary systems.

The numerous, though fragmentary studies so far make it very difficult to predict the three-dimensional evolution of planets and their interaction with their host discs. In answering the question concerning the evolution of a circumstellar disc, one could make a major step in the discussions about the formation of misaligned Hot Jupiters.

 In view of the above discussion about misaligned planets, it is important to investigate the evolution of a protoplanetary disc experiencing gravitational perturbations of a stellar flyby event as previous studies \citep[e.g.][]{Xia2014} have shown that a gaseous circumstellar disc is able to control the evolution of the whole disc-planet system. 
In this paper, we study the three-dimensional evolution of a viscous protoplanetary disc which is perturbed by a passing star. The aim is to answer the question whether single stellar flybys are capable to generate significant total disc inclinations. 
Furthermore, by surveying the parameter space of stellar close encounters we aim to find specific parameter domains which allow inclination generation of the disc. 
We make the simplification of neglecting the disc self-gravity which results in a significant reduction in computational resource requirements. This is mainly motivated by the fact that disc masses are usually very small compared to the central stellar mass.

In studying the disc inclination, we note that warping of protoplanetary discs can generally occur as a result of tidal interaction with a binary star or if the disc accretes matter with angular momentum misaligned with that of the star \citep[e.g.][]{Ter1993, Pap1995, Bat2010}. For stellar perturbers on parabolic orbits, the extent of warping for different orbital parameters is still an open question due to lack of three dimensional studies. However the potential warping of a protoplanetary disc is relevant in the discussion of planets since it has direct consequences on the formation and three-dimensional evolution of planets in these systems \citep[see also][]{Ter2013}.

It has been shown that smoothed particle hydrodynamics (SPH) simulations are capable of being applied to the problem of three-dimensional evolutions of gaseous discs rather than grid-based simulation methods. In contrast to grid-based methods, as they adopt a Lagrangian approach, SPH simulations can be readily used to simulate a gas disc  with a free boundary   that undergoes a large amount of movement  in three dimensions with the disc having the freedom to change its shape at will. As it is suitable for the system that we aim to study, we adopt this approach  here. 
In Section \ref{sec:sim_details}, we describe our simulation technique, giving details  of the equation of state (eos), the smoothing length and artificial viscosity.
In Section \ref{sec:IC}, we describe the general setup for  the disc and the central star.
 For a better understanding of the results, we discuss the properties of realistic close encounters in more detail in Section \ref{sec:orbit}.
We go on to present the simulation results in  Section \ref{sec:results}.
Finally, we summarize and discuss our results in Section \ref{sec:conclusions}.

\section{Simulation details}\label{sec:sim_details}

We  have performed simulations using a modified version of the publicly available code {\rm GADGET}-2  \citet{Spr2005}. 
GADGET-2 is a hybrid N-body/SPH code capable of modelling both fluid and distinct  fixed or  orbiting  massive bodies. In our case the central star, of mass $M_*,$  is fixed while the stellar perturber, of mass $M_p$ moves as a massive body.
We adopt spherical polar coordinates $(r, \theta,\phi)$ with origin at the centre of mass of the central star.  The associated Cartesian coordinates $(x,y,z)$ are such that the $(x,y)$ plane coincides with the initial midplane of the disc.

\noindent  The gaseous disc is represented by gasesous (SPH) particles. 
An important issue  for N-body/SPH simulations is the choice of the gravitational softening lengths. The only gravitational softening that is included in our simulation is for the gravitational interaction between the SPH particles and the two stellar particles.

The total unsoftened gravitational potential $\Psi$ at a position ${\bf  r}$ is given by \citep[see also][]{Pap1995, Lar1996}.
\begin{eqnarray}
 \Psi({\bf  r})&=& -\frac{G M_* }{|{\bf  r}|} - \frac{G M_p }{|{\bf  r}-{\bf  D}|} + \frac{G M_p {\bf  r}\cdot {\bf D} }{|{\bf  D}|^3} \ .
\label{eq:Pot}\end{eqnarray}
Contributions from the central star and the stellar perturber of mass, $M_p,$ with  position  ${{\bf  D}}$, are included.
Because  the origin of the coordinate system  moves with  the central star, the  well known  indirect terms accounting for the acceleration of the coordinate system are present in equation (\ref{eq:Pot}).

For the  computation  of the gravitational interaction between the
massive bodies  and the gas particles,  the potential was softened following the   method of \citet{Spr2005}.
This was implemented with  fixed  softening lengths $\varepsilon_*=\varepsilon_p=0.01$ internal length units for both the central star and the stellar perturber.
The disc's self-gravity was neglected. 
The stars are allowed to  accrete gas particles that approach them very closely. We  followed  the procedure  of \citet{Bat1995}. 
This was applied such that for the stars, the outer accretion radius was fixed during the simulation to be $R_a=0.02$ internal length units. The inner accretion radius was taken to be 0.5 of the outer accretion radius.

We adopt a locally isothermal equation of state (EOS) 
\begin{eqnarray}
c_s = h |{\bf  v}_\varphi| \ . 
\end{eqnarray}
Here, $h=H/r$ is the disc aspect ratio  with $H$  being the circumstellar  disc scale height and $r=|{\bf  r}|$ being the distance to the central star. The rotational velocity vector in the circumstellar disc is  ${\bf  v}_\varphi$. As the disc is perturbed by a star on an inclined orbit, its initial orientation can change significantly. In order to find the exact direction of the rotational velocity vector, we apply the general description for ${\bf  v}_\varphi$ by using the following relations in three dimensions:
\begin{equation}
 {\bf  v}_\varphi={\bf  \omega} \times {\bf   r} \ , \\
 {\bf  \omega}=\frac{{\bf  r} \times {\bf  v}}{r^2}\ .
 \end{equation}
${\bf  v}$ is the velocity vector and ${\bf  \omega}$ the angular velocity vector of a gas particle.

For  $r \leq 4$, we adopt  $h = 0.05$. For previous studies with a binary companion on an orbit inclined to the disc, it was found that  for high initial inclinations $\ge 60\degr$, expansion of the outer regions of the disc could result in a small number of particles  being induced to interact strongly with the companion. In order to avoid numerical problems arising from this, we adopted the following procedure. For the outer parts with  $5 > r >4$, we applied a 
linear decrease of $h$ to $0.03$. For $r\geq 5$, we adopted  a constant ratio $h = 0.03$.

In addition, we include an routine to allow gas particles that become bound to the perturber to have an appropriate hydrodynamical treatment. In these cases, the hydrodynamical parameters are computed by taking the distance of a gas particle to the perturber instead of its distance to the primary star.
We found that the application of this procedure did not affect the global outcomes of the simulations.

\subsection{Smoothing length and artificial viscosity} \label{sec:artv}

For our SPH calculations, the smoothing length was adjusted so that the number of  nearest  neighbours  to any particle contained within a sphere of radius equal to the local smoothing length was  $40 \pm 5$. 
The pressure is given by  $p=\rho c_s^2$. 
Thus, the temperature in the disc is $\propto r^{-1}$. The artificial viscosity parameter $\alpha$ of GADGET-2 \citep[see equations (9) and (14) of][]{Spr2005} was taken to be $\alpha =0.5$. Note that this $\alpha$ parameter is not the usually applied $\alpha$-parameter in the context of $\alpha$-viscosity and not the $\alpha$ parameter for artificial viscosity in most other SPH simulation codes either. A detailed discussion of our applied $\alpha$ parameter can be found in \citet{Spr2005}. 

We remark that the artificial viscosity is modified by the application of a viscosity-limiter to reduce artificially induced  angular momentum transport in the presence of shear flows. This is especially important for the study of Keplerian discs.
Details are given in \citet{Xia2013} where we showed that runs with  $\alpha=0.5$  provided the best match to the analytic ring spreading solution with the \citet{Sha1973} viscosity parameter  $\alpha_{SS}=0.02.$
 But note that this correlation with $\alpha_{SS}=0.02$ is only valid for Keplerian thin discs without significant vertical expansion of the disc.

\section[]{Initial conditions} \label{sec:IC}

We study a system composed of a central star of one solar mass $M_\odot$, a gaseous disc surrounding the central star and a stellar perturber of
mass  $M_p$.
The disc is set up such that the angular momentum vector for all particles was in the same direction enabling a midplane for the disc to be defined.

The primary central star is  fixed at the origin of our coordinate system while the perturber follows a parabolic orbit about the primary central star. 
The gravitational effect of the disc on the perturber's orbit is neglected. In order to  avoid transients arising from an abrupt introduction of the perturbing star,  we allow its mass  to increase linearly to $M_p$ during the first 5 internal time units, with the internal time unit  being the orbital period at $a=5\ \mathrm{au}$. Corresponding to this, the internal unit of length is taken to be $5\ \mathrm{au}$. These units of length and  time are  adopted  for all plots shown in this paper.
But note that for given sufficient resolution in the disc, our results can be generalized to larger disc radii since the problem studied here is scale-invariant. As we will discuss in the next section, the perturber and its orbit can be expressed as functions of the central star and the initial disc size making it possible to apply the scenario on different disc sizes.


The particle distribution  was chosen to  model  a  disc with surface density profile  given by
\begin{eqnarray}
\Sigma=\Sigma_0 R^{-1/2}. \label{eq:sigma}
\end{eqnarray}
Here $\Sigma_0$ is a constant and $R$ is the radial coordinate of a point in the midplane.
This applied to  the radial domain $[0, R_{\rmn{out}}-\delta]$,  $R_{\rmn{out}}= 5a$ and $\delta =0.4 a$. 
  A taper was applied  such that the surface density was set to decrease linearly to zero
  for $R$ in the  interval $[ R_{\rmn{out}}-\delta,  R_{\rmn{out}}+\delta].$
Disc  material thus  occupies the radial domain  $[0, R_{\rmn{out}}+\delta]$. 
 
\noindent The disc mass is given by
\begin{eqnarray}
M_D=2\pi \int_{R_{\rmn{in}}}^{R_{\rmn{out}}} \Sigma(r)r dr=\frac{4}{3} \pi \Sigma_0 R_{\rmn{out}}^{3/2}\ ,
\end{eqnarray}
which is used to determine $\Sigma_0.$
For the simulations presented here, we adopted $M_D = 10^{-2}\ \rmn{M}_*.$
 As self-gravity is expected to play a minor role in such a low-mass disc, it is neglected in the simulations.

The disc particles were set up in a state of pure Keplerian rotation according to
\begin{eqnarray}
 v_\varphi=\sqrt{r\frac{d\Psi_*}{dr}}\ ,
\end{eqnarray}
where $\Psi_*$ is the gravitational potential due to the central star.
In the innermost region around the central star, the disc properties (e.g. $v_\varphi$ and accordingly $c_s$) are modified by  the  softening of the potential due to the central star.


The  number of gas particles   for most of  our simulations was taken to be  2$\times 10^5.$
This enabled a large suite of simulations to be carried out. 
Simulations were stopped when the disc inclinations and disc masses became constant after the stellar encounter. 
Some of our simulations were also run with $4\times 10^5$ particles in order to test the effect of changing particle number. No significant changes occured. 

In this paper, we will mainly concentrate on the evolution of a viscous gaseous disc which experiences a single stellar flyby. Due to numerical reasons in the hydrodynamical simulations, the long-term evolution of the disc cannot be followed accurately as we will discuss in Section \ref{sec:longterm} .
For comparison, we have also performed a set of simulations with a disc composed of collisionless particles with a purely N-body treatment.

\section{Properties of realistic close encounters} \label{sec:orbit}

\subsection{Scale invariance}

Although the specific choice of simulation parameters relating to the initial disc and the perturber's orbit were chosen largely for numerical convenience, we point out that the results can be extended to apply to other configurations. 

In the system that we aim to study, we are able to apply the usual scaling of lengths and times that leave gravitationally interacting systems invariant.
More precisely, the numerical outcome is expected to scale with the ratio of  the pericenter to the disc's outer radius. For systems with different disc's radii, the results are identical as long as the pericenter is chosen such that its ratio to the disc's radius is unchanged and if the time is scaled accordingly.

In our study, the outcome is expected to depend on the stellar mass ratio. However, we do not expect a simple dependency of the final disc properties such as disc inclination on the mass ratio. This is because the disc is allowed to lose mass to the perturber and thus the number of disc particles that contribute to the disc properties can differ significantly between distinct simulation models.

Beside the stellar mass ratio, the possible effect of different three-dimensional orientations of the perturber's orbit on the final disc properties is a major issue that has not been studied yet and thus has to be investigated. In section \ref{sec:3dorbit}, we will elaborately discuss the characterization of the orbit in our simulations.

\subsection{Stellar mass ratios} \label{sec:mass}

In stellar clusters, stellar masses follow a global distribution function which is called the initial mass function (IMF) \citep[e.g.][]{Kro2002, Cha2003}. Taking into account the IMF in a stellar cluster, it can be very common that close encounters occur for different mass stars. 
Several numerical surveys have been performed to study the frequency of close encounters in different cluster types \citep[e.g.][]{Olc2012, Pfa2013, Vin2015}. 
\citet{Pfa2013} studied the encounter history and characteristics of different stellar clusters. In their work, they determined the probability of close encounters with the help of  N-body simulations. 
For example, Figure 6 of \citet{Pfa2013} shows the probability distribution of close encounters for different perturber masses influencing solar-like stars. 
At $t=2.8\ \rmn{Myr}$, the probability distribution shows two local maxima, one for low-mass perturbers and another one for massive perturbers with masses considerably higher than 1 $\rmn{M_\odot}$. 
As we are interested in close encounters with perturber masses $\geq 5\ \rmn{M_\odot}$, it is important to determine the probability of a solar type star to have an encounter with a perturber with at least 5 $\rmn{M_\odot}$ in order to confirm the likelihood of such a scenario. This required probability is the sum over all probabilities in the range [5: *] $\rmn{M_\odot}$ in  Figure 6 of \citet{Pfa2013}. 
Given the numerical results for $t=2.8\ \rmn{Myr}$, we determined the probability for a solar-like star to have had at least one encounter with a perturber with at least $5\ \rmn{M_\odot}$ between 100 au and 1000 au to be roughly 1/4. Thus, every fourth solar-like star is expected to have had at least one encounter with a massive perturber during the first 2.8 Myr of its lifetime.

As our results are length invariant, this relatively high propability also applies to our simuliations with smaller disc sizes. 

So far, numerical cluster simulations have not studied the likelihood of different three-dimensional orientations of the perturber's orbit relative to the initial disc plane making it difficult for us to make predictions of the probability of a stellar flyby with a specific orientation.

For this reason, we assume that the relative orientation between a perturber's orbit and a disc in a stellar cluster is randomn. Hence, we could assume that prograde and retrograde orbits are equiprobable. Applying this assumption, the probability of a solar-like star to have at least one close encounter on a retrograde orbit is expected to be $\sim$ 0.125. 

Given this relatively high probability of close encounters between massive stars on retrograde orbits and solar-like stars, 
it would be interesting to study their effects on the three-dimensional evolution of circumstellar discs in order to better understand the possible formation of misaligned planets. 
Thus, we will perform simulations with perturber masses in the range [5:100].

\subsection{Perturber's orbit} \label{sec:3dorbit}

\begin{figure}
\centering
\includegraphics[width=5cm]{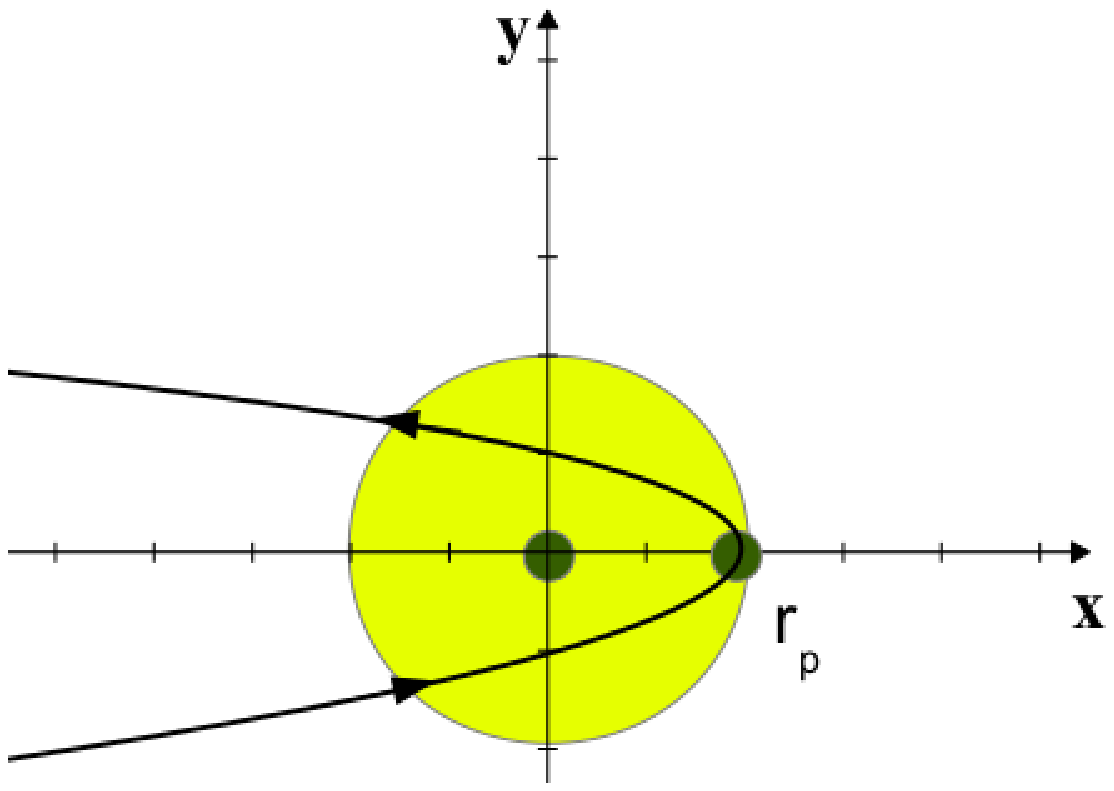} 
\includegraphics[width=5cm]{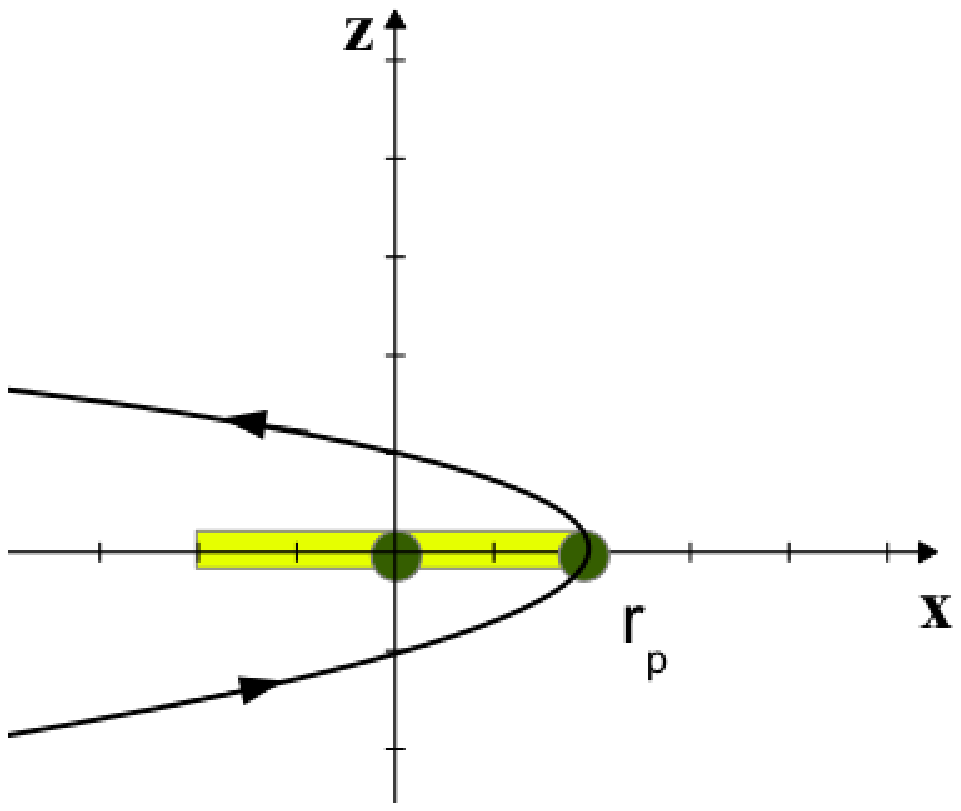}
\includegraphics[width=4cm]{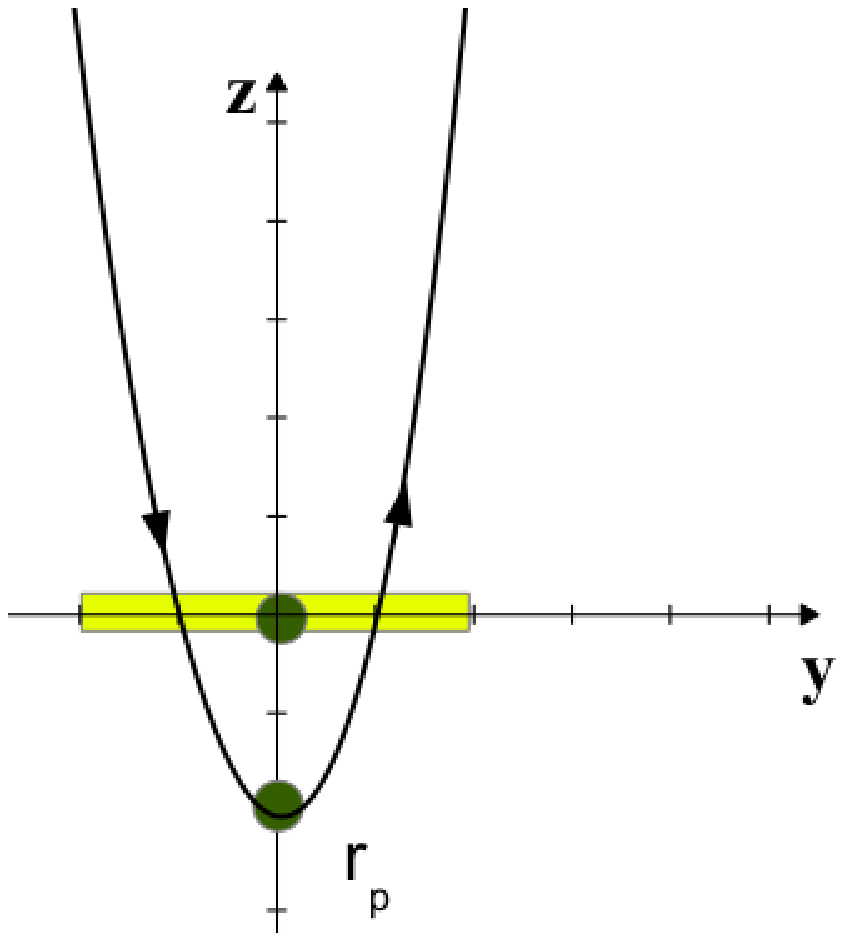}
\caption{Upper panel: Orbit of the stellar perturber in the non-rotated case with the orbit being in the $(x,y)$-plane. Middle panel: rotation about the $x$ axis by $\beta_p=90\degr$ while $\gamma=0\degr$. After this rotation, the stellar orbit is in the $(x,z)$-plane. In this picture, the $y$-axis points away from the reader. Lower panel: Rotating the initial stellar orbit about the $y$ axis by $\gamma_p=90\degr$ (while $\beta_p=0\degr$) leads to a stellar orbit in the $(y,z)$-plane. Here, the $x$-axis points towards the reader. In all pictures, the yellow areas show the location of the gaseous disc.  The green filled circles mark the two stars at pericenter passage. }
\label{fig:orbit}
\end{figure}

The present study focuses on stellar perturbers on parabolic orbits ($e=1$). Previous studies have shown that even for ONC-like clusters, most encounters are expected to be close to parabolic \citep[e.g.][]{Lar1990, Ost1994, Olc2010, Pfa2013}. 

The relative orbit of the stellar perturber to the central star is a parabolic orbit which can be rotated in space. We define the pericenter of the orbit in the $(x,y)$ plane (see also uppermost panel in Fig. \ref{fig:orbit}). The orbital plane of the perturber is then rotated about the $x$ axis by the angle $\beta_p$ (anti-clockwise) and the $y$ axis by the angle $\gamma_p$ (clockwise).

Fig. \ref{fig:orbit} shows the original stellar orbit in the $(x,y)$-plane (upper panel) and two extreme cases of ($\beta_p=90\degr, \gamma_p=0\degr)$ (middle) and ($\beta_p=0\degr, \gamma_p=90\degr)$ (lower panel). The green filled circles mark the location of the central star and the pericenter of the stellar perturber's orbit. The large yellow filled circle in the upper panel and the yellow bulks in the two lower panels show the orientation of the protoplanetary disc.

 As one can see clearly, a rotation about the $x$ axis does not change the location of the pericenter. At the pericenter, the stellar perturber crosses the disc. In contrast to the rotation about the $x$ axis, the lowest panel shows that when rotating the initial stellar orbit about the $y$ axis, the pericenter is moved out of the gas disc. In this case, the stellar perturber crosses the disc twice.

The location of the pericenter has a significant meaning with respect to the torque that the perturber acts on the disc.
As \citet{Kor1995} have shown, most of the torque experienced by the disc is applied at the point of pericenter passage. 
For small perturbations, there are analytical studies of the influence of a perturber on a gaseous disc.
For example, \citet{Ost1994} applied the linear perturbation theory on an inviscid hydrodynamical model in order to derive an asymptotic expression for the angular momentum loss of the disc to the encounter. \citet{Kor1995} in contrast used a Fourier expansion in azimuthal modes and were able to calculate the angular momentum exchange without asymptotic assumptions. 
Finally, \citet{Lar1997} performed a simplified analysis with the most significant parts of the disc response which have been identified by the previous works and obtained expressions for the angular momentum exchange for coplanar as well as inclined parabolic orbits of the perturber. 
However, for stellar flybys where the perturbation of the passing star is significantly large, the disc is strongly perturbed and its global evolution is disrupted such that these analytical estimates are not applicable.

\subsection{Parameter space of the perturber} \label{sec:parameterspace}

Every simulation is characterized by the stellar mass ratio $P_m=M_p/M_*$, the impact parameter (periastron distance relative to the outer disc radius) $q=r_p/R_{\rmn{out}}$ and the two angles $\beta_p$ and $\gamma_p$.  
The orbit of the perturber is computed such that the perturber passes its pericenter at $t=50$ internal time units. In order to cover the parameter space as completely as possible, we perform surveys in the following ranges: $P_m=$[1, 5, 10, 20, 50, 100], $q=$[1, 2, 5, 10], $\beta_p$=[0, 45, 90, 135, 180]$\degr$, $\gamma_p$=[0, 45, 90, 135, 180]$\degr$.
We will first show results of the two extreme cases of $(\beta_p=0\degr$, $\gamma_p\neq 0\degr)$ and $(\beta_p\neq 0\degr, \gamma_p=0\degr)$ and then proceed with surveys with both $\beta_p$ and $\gamma_p$ being $\neq 0\degr$.

\section{Simulation results}\label{sec:results}

\subsection{$q=1$, $\gamma_p=0\degr$, survey of $\beta_p$} 

\begin{figure}
 \centering
\includegraphics[width=8cm]{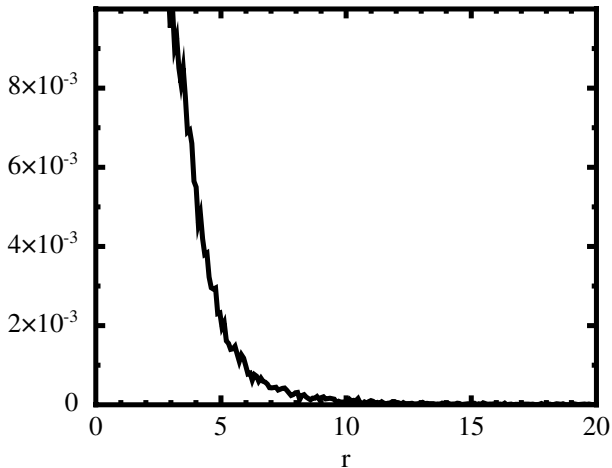}
\caption{The azimuthally averaged disc surface mass density profile (in units of $M_\rmn{J}/(5\rmn{au})^2$) at 150 internal time units as a function of distance (in units of 5 au) for a run with $\gamma_p=0\degr$, $\beta_p=135\degr$, $P_m=20$ and $q=1$.  }
\label{fig:mass_distribution}
\end{figure}

Figure \ref{fig:mass_distribution} shows the azimuthally averaged disc surface mass density profile (in units of $M_\rmn{J}/(5\rmn{au})^2$) as a function of distance (in units of 5 au) for a run with $\gamma_p=0\degr$, $\beta_p=135\degr$, $P_m=20$ and $q=1$ at a later stage after the stellar flyby. As can be seen clearly, the mass outside $r=10$ internal  length units tends to 0 and therefore will be ignored in the following. For the determination of the final total disc mass, disc angular momentum and disc inclination, we will only consider gas particles inside 10 length units.

\begin{figure}
 \centering
\includegraphics[width=9cm]{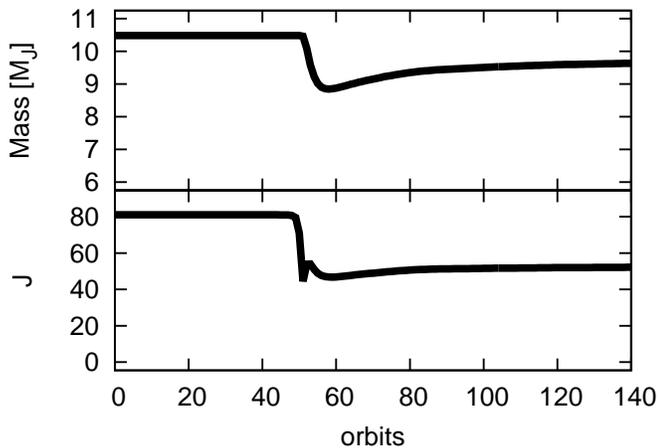}
\caption{Evolution of total gas mass (upper panel) and disc angular momentum  (lower panels) inside 10 length units for $\gamma_p=0\degr$, $\beta_p=135\degr$, $P_m=20$ and $q=1$.  }
\label{fig:rp_1,gamma_0,example_1}
\end{figure}

Fig. \ref{fig:rp_1,gamma_0,example_1} shows the evolution of the total gas mass (upper panel) and disc angular momentum (lower panel) for particles inside 10 length units for the same simulation as shown in Fig. \ref{fig:mass_distribution} ($\gamma_p=0\degr$, $\beta_p=135\degr$, $P_m=20$ and $q=1$). 
The total angular momentum of the disc is defined as 
\begin{eqnarray}
|{\bf J}_D|=\left| \sum_i m_i ({\bf r}_i \times {\bf v}_i) \right|  \ . \label{eq:J}
\end{eqnarray}
${\bf J}_{D}$ is the total angular momentum vector of the disc. The summation in (\ref{eq:J}) is  taken over all active gas particles which are located inside 10 length units and bound to the primary star.

\begin{figure}
 \centering
\includegraphics[width=9cm]{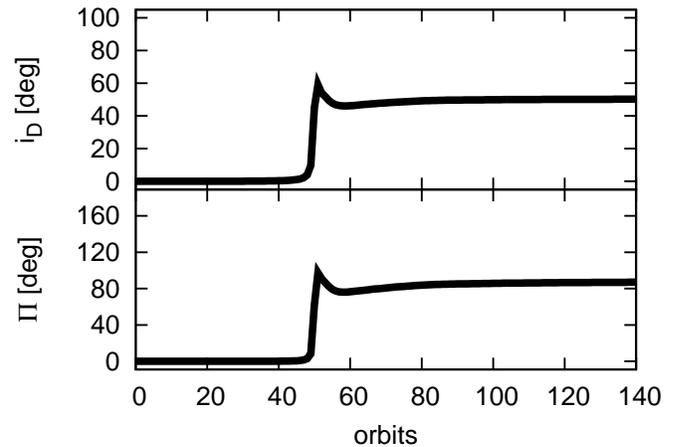}
\caption{Evolution of the total disc inclination with respect to the $(x,y)$ plane (upper panel) and the precession angle (lower panel) for $\gamma_p=0\degr$, $\beta_p=135\degr$, $P_m=20$ and $q=1$.  }
\label{fig:rp_1,gamma_0,example_2}
\end{figure}

Fig. \ref{fig:rp_1,gamma_0,example_2} shows the evolution of the total disc inclination $i_D$ with respect to the $(x,y)$ plane (upper panel) and the precession angle (lower panel) for the same simulation setup as shown in Fig. \ref{fig:mass_distribution}.
The disc inclination $i_D$  is obtained from
\begin{eqnarray}
\hspace{-1mm}&i_{D}&= \arccos\left( \frac{ J_{D,z} }{|{\bf J}_{D}|} \right)\ , 
\end{eqnarray}
where $J_{D,z}$ is the $z$-component of the disc angular momentum vector ${\bf J}_D$.

The precession angle $\Pi$ of the disc is defined as
\begin{eqnarray}
\cos \Pi &= &\frac{{\bf J}_D \times {\bf J}_p}{|{\bf J}_D \times {\bf J}_p|} \cdot {\bf u}\, ,
\end{eqnarray}
with ${\bf J}_p$ being the orbital angular momentum vector of the perturber 
 \citep[see also][]{Xia2014}. For ${\bf u}$, any fixed unit reference vector in the orbital plane of the perturber could be used.
 In our case, we made  the following choice:
\begin{eqnarray}
 {\bf u}=\frac{{\bf J}_{D,0} \times {\bf J}_p}{|{\bf J}_{D,0} \times {\bf J}_p|}\ ,
\end{eqnarray}
where  ${\bf J}_{D,0}$ is the total disc angular momentum vector at time $t=0$.  

During the stellar flyby, retrograde precession of the disc angular momentum vector ${\bf J}_D$ about the orbital angular momentum vector of the perturber ${\bf J}_p$ is expected to occur \citep[see also][]{Lar1997}. The strongest effect of the perturber occurs at its pericenter passage which results in a considerable increase of the precession angle as can be found in the lower panel of Fig. \ref{fig:rp_1,gamma_0,example_2}.

\begin{figure}
 \centering
\includegraphics[width=9cm]{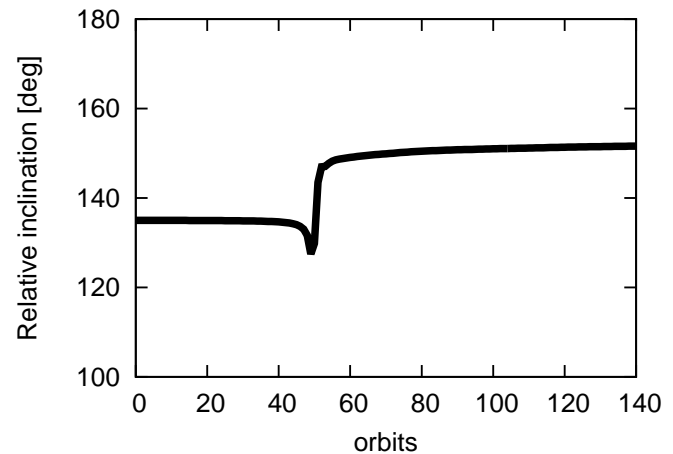}
\caption{Relative inclination of the disc angular momentum vector to the orbital angular momentum vector of the perturber as a function of time for $\gamma_p=0\degr$, $\beta_p=135\degr$, $P_m=20$ and $q=1$.  }
\label{fig:rp_1,gamma_0,example_3}
\end{figure}

In addition to disc precession, \citet{Pap1995, Lar1997} derived the inclination evolution of the disc due to the $y$-component of the tidal torque. For small perturbations to which the disc response is linear, they concluded that the final orientation of the disc is mainly determined by the precession rather than the inclination changes. But as the disc response in our simulations is in the non-linear regime, the relative importance of inclination evolution compared to precessional effects has to be investigated anew.

For this, Fig. \ref{fig:rp_1,gamma_0,example_3} shows the time-dependent evolution of the relative inclination of the disc angular momentum vector to the orbital angular momentum vector of the perturber for the same simulation setup as shown in Fig. \ref{fig:mass_distribution}.
The relative inclination $i_{rel}$ is defined as
\begin{eqnarray}
 i_{rel}= \arccos\left( {\frac{{\bf J}_{D} \cdot {\bf J}_{p}}{|{\bf J}_D| \cdot |{\bf J}_p|}} \right)\ . \label{eq:i_rel}
\end{eqnarray}

In cases where the precessional effects mainly dominate the final orientation of the disc, the relative inclination between the disc and the perturber's orbit is expected to be constant. However, Fig. \ref{fig:rp_1,gamma_0,example_3} demonstrates clearly that in our simulation, the relative inclination evolves significantly by $\sim 15^\circ$. As a consequence, in the example simulation shown by Fig. \ref{fig:mass_distribution} - \ref{fig:rp_1,gamma_0,example_3}, the final orientation of the disc is  determined by both the disc orbital inclination change and precession of the disc.



Shortly after the pericenter passage of the perturber at $t=50$ orbits, the system relaxes quickly and forms a constant final configuration at $t\gtrsim 100$ orbits. From this time on, all parameters of interest stay mainly constant.
Since all simulations follow a similar time-dependent evolution and as we are only interested in the final outcome of the remaining gaseous disc, 
we will only show the final values for the different parameters such as disc mass or disc inclination at the end of the simulations.

\begin{figure}
\centering
\includegraphics[width=8cm]{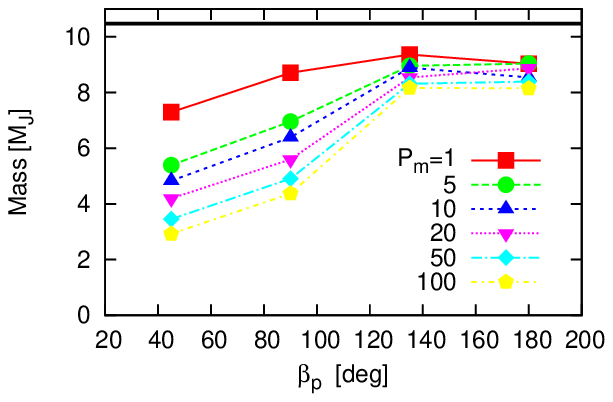}
\includegraphics[width=8cm]{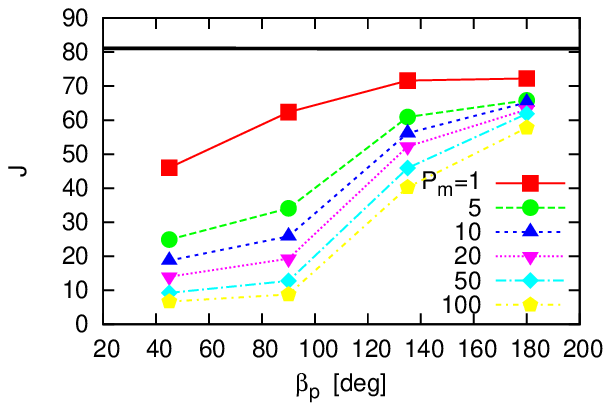}
\caption{Total gas mass (upper panel) and angular momentum (lower panel) inside 10 length units as function of the angle $\beta_p$ for stellar mass ratios in the range [1: 100] as indicated by the different colors. For all simulations $\gamma_p=0\degr$. The black horizontal lines indicate the total mass and angular momentum of the initial disc.}
\label{fig:rp_1,gamma_0,m,l}
\end{figure}

Figure \ref{fig:rp_1,gamma_0,m,l} shows the final disc mass (upper panel) and angular momentum (lower panel) 
for a stellar flyby at $q=1$, $\gamma_p=0\degr$ and different values of $\beta_p$. Note that $\beta_p<90\degr$ represents the prograde orbit while $90\degr<\beta_p<180\degr$ represents the retrograde case.

Stellar mass ratios in the range [1: 100] are considered. The simulations with $\gamma_p=0\degr$ and $\beta_p \neq 0\degr$ represent the case shown in the middle picture of Figure \ref{fig:orbit}. In this case, the pericenter of the parabolic orbit is still located inside the initial disc's plane.

For a given $P_m$, the disc mass loss decreases with increasing $\beta_p$. The remaining disc angular momentum after the encounter increases with increasing $\beta_p$. This is in agreement with previous studies. 
The prograde encounters ($\beta_p < 90\degr$) are more destructive \citep[see also][]{Too1972, Cla1993} than the retrograde ones ($90\degr < \beta_p < 180\degr$). 
In the prograde case $\beta_p=45\degr$, more than 50\% of the initial disc mass is lost to the stellar perturber for $P_m>20$. For $\beta_p=90\degr$, $P_m>50$ leads to more than 50\% mass loss.
For these high $P_m$ ratios, the more massive star is generally expected to have its own gaseous envelope. Hence, the results shown for higher mass ratios $P_m$ are only partly applicable to realistic encounters. Nonetheless, it is important to study the dependency of the disc evolution for different values of $P_m$. 

For increasing $P_m$ which is associated with an increased gravitational perturbation on the disc, the disc mass loss as well as the removal of angular momentum increases significantly.
This trend is a common outcome for all simulations. Although there are few differences in the numerical simulation methods and studies of disc mass loss betwen our work and previous papers \citep{Pfa2005, Olc2006}, some parameter sets allow a qualitative comparison between the simulation results.
\citet{Pfa2005} have performed a parameter study of star-disc encounters by treating the disc particles as test particles without self-gravity or viscous forces.
They have studied different stellar mass ratios and inclined orbits and the resulting disc mass loss and angular momentum loss.
For the retrograde case with $\beta_p=180\degr$, $P_m=1$ and $q=1$, \citet{Pfa2005} found a relative disc mass loss of $0.15 \pm 0.04$ and relative angular momentum loss of $0.17 \pm 0.02$ \citep[see Figure 4 and 5 of ][]{Pfa2005}.  In our case, the relative mass loss is $\sim 0.05$ and the relative angular momentum loss is $\sim 0.13$. The slight differences in the mass loss and angular momemtum loss can be explained by the viscous forces that operate in our simulations which in general result in less disc mass loss and hence also less angular momentum loss. Taking into account the different numerical descriptions of the discs, the results show satisfactory agreement with each other. 
In Figure 7 of \citet{Pfa2005}, the disc mass loss as well as angular momentum loss decreases with increasing inclination angle. This is also the general outcome of our survey shown in Figure \ref{fig:rp_1,gamma_0,m,l}.
The general outcome is a more destructive nature of prograde (inclined) orbits than retrograde ones. Thereby, the disc mass loss is closely related to the stellar mass ratio $P_m$ and the impact parameter $q$. The disc mass loss and angular momentum loss increase for decreasing $q$ and increasing $P_m$.
All dependencies can be found in our simulations as well as in those of \citet{Pfa2005} and the results show approximate agreement.

\begin{figure}
\centering
\includegraphics[width=8cm]{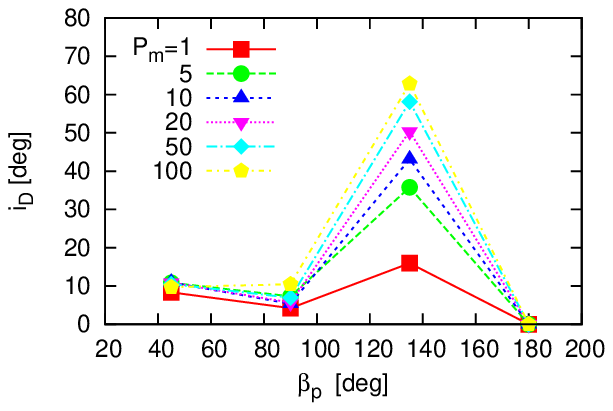} 
\includegraphics[width=8cm]{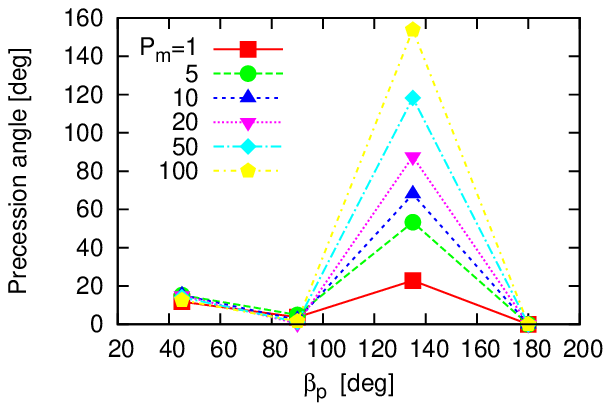} 
\includegraphics[width=8cm]{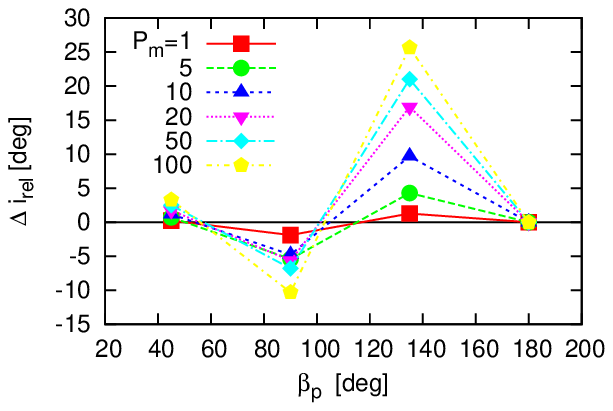} 
\caption{ Disc inclination $i_D$ with respect to the $(x,y)$ plane (upper panel), precession angle (middle panel) and change of relative inclination (lower panel), as a function of $\beta_p$ for stellar mass ratios in the range [1: 100] as indicated by the different colored points. For all simulations $\gamma_p=0\degr$. }
\label{fig:rp_1,gamma_0,i}
\end{figure}

Figure \ref{fig:rp_1,gamma_0,i} shows the final disc inclination $i_D$ with respect to the $(x,y)$ plane (upper panel), the precession angle (middle panel) and the change of the relative inclination (lower panel) for the same simulations as shown in Figure \ref{fig:rp_1,gamma_0,m,l}.
The change of the relative inclination is defined as
\begin{eqnarray}
\Delta i_{rel}= i_{rel} - i_{rel,0}\ .
\end{eqnarray}
$i_{rel,0}$ is the initial relative inclination at time $t=0$ and $i_{rel}$ the final relative inclination between the disc angular momentum vector and the perturber's orbital angular momentum vector. For the parameter set ($\gamma_p=0$ and $\beta_p \neq 0$), the initial relative inclination is $i_{rel,0}=\beta_p$.

Comparing the different simulations, the case of $\beta_p=180\degr$ shows the lowest disc inclination generation. For all values in the range $P_m = [1: 100]$, the disc inclination is approximately 0. 
For $\beta_p=90\degr$, the stellar perturber is only able to generate disc inclinations below $10\degr$. In this case, precession effects are mainly absent. It is rather the change of the disc orbital inclination which determines the final orientation of the disc.
In the prograde case with $\beta_p=45\degr$, $P_m$ does not have a significant effect on the final disc inclination. The final disc inclination is $i_D \cong 10\degr$ for all mass ratios.

The largest influence of the perturber can be found for the retrograde case of $\beta_p=135\degr$. In this case, the  final disc inclination is closely correlated to the mass ratio $P_m$. Significant disc inclinations can be generated up to $60\degr$ for the two extreme cases with $P_m=50$ and $P_m=100$. For $P_m=1$, the final disc inclination is only 15\degr. However, the increase of $P_m$ to 5 shows a final disc inclination of more than 35\degr. For this reason, equal-mass stars with $P_m=1$ seem to represent an extreme case of very low disc inclination generation whereas a slight increase of $P_m$ can already lead to significant disc inclination growth. 
In this retrograde inclined case, both the precession angle and the change of relative inclination increase with increasing $P_m$. The significant changes of the relative inclination imply that the final disc orientation in these cases is not only determined by the precession of the disc angular momentum vector about ${\bf J}_p$, but rather significantly affected by the change of the disc orbital inclination.

Taking together the results for the disc inclination and the disc mass loss, the results can be interpreted in the following way.
In the prograde case, gas particles which are strongly affected by the perturber, are mostly removed by the perturber. Thus, they cannot contribute to the disc inclination. Particles which are still left in the disc, only have minor orbital inclinations. In total, the disc inclination is only marginal. 
The high final disc inclinations in the retrograde case are produced by a large fraction of remaing disc material. In these cases, the highly inclined gas particles are not removed by the perturber and thus can contribute to the total disc inclination.  The final total disc inclination is a result of a combination of the disc orbital inclination change and precession about the orbital angular momentum vector of the perturber. 

For $\gamma_p=0\degr$, we conclude that prograde orbits generally do not allow significant disc inclinations because of the significant disc mass loss. Simulations with $P_m$ up to 100 failed to produce any significant disc inclinations above $10\degr$. The most promising cases for the generation of disc inclination are retrograde inclined parabolic orbits. Significant disc inclinations up to 60\degr can be generated for $90\degr<\beta_p<180\degr$ and mass ratios $5\leq P_m \leq 100$.


\begin{figure} 
\centering
\includegraphics[width=8cm]{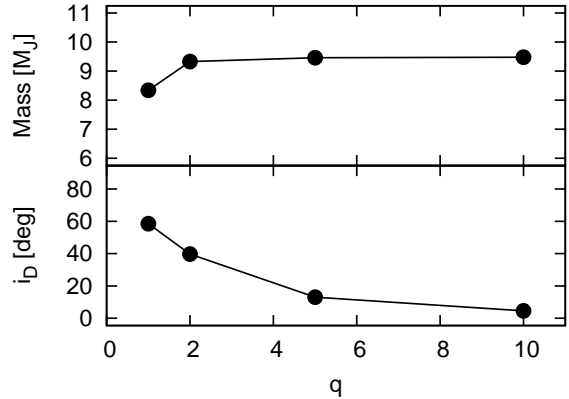}
\caption{ Final disc inclination with respect to the $(x,y)$ plane (lower panel) and total disc mass (upper panel) for $P_m=50$. The rotation angles are $\gamma=0\degr$ and $\beta_p=135\degr$.  }
\label{fig:q}
\end{figure}

\subsubsection{Dependence on $q$ }


The gravitational influence of the stellar perturber is expected to depend on its distance to the disc which can be  studied by varying the impact parameter $q$.
Fig. \ref{fig:q} shows the final disc inclination (lower panel) and total disc mass (upper panel) for the case $\gamma=0\degr$, $\beta_p=135\degr$, $P_m=50$ and different values of $q$.

It is generally expected that close stellar encounters with impact parameters $q\leq 3$ produce significant changes in the global structure of the gaseous disc \citep[e.g.][]{Ost1994}.
This expectation is confirmed in Figure \ref{fig:q}. For $q<3$, the stellar perturber is capable to generate significantly large disc inclinations while for larger values of $q$, the final disc inclination is $\lesssim 13\degr$.

At the same time, the disc mass loss decreases with increasing $q$. While for $q\leq 2$, the disc mass loss is visible, the cases with $\geq 5$ do not show significant disc mass. 

The above results for $\gamma_p=0\degr$ lead to the conclusion that significant disc inclinations can be generated for retrograde inclined cases with $90\degr<\beta_p<180\degr$, impact parameters $q\leq 3$ and stellar mass ratios $P_m\geq 5$. 
In all other cases 
final disc inclinations above 20\degr cannot be achieved.

\subsection{$q=1$, $\beta_p=0\degr$, survey of $\gamma_p$ } 

\begin{figure}
\centering
\includegraphics[width=8cm]{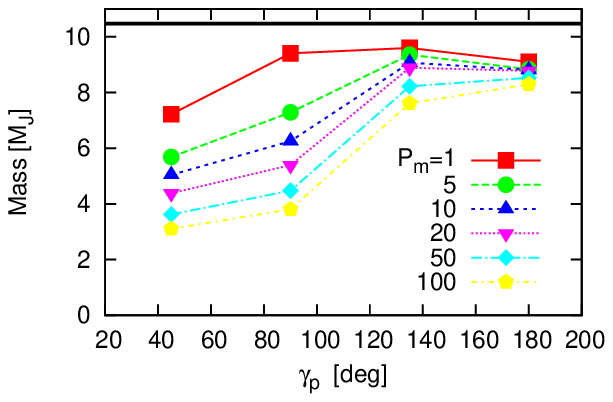}
\includegraphics[width=8cm]{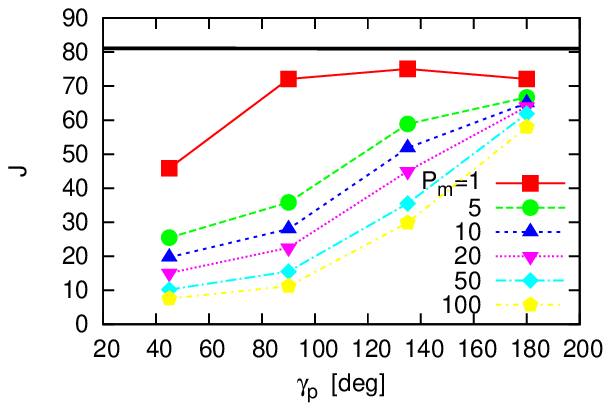}
\caption{Total gas mass (upper panel) and angular momentum $|{\bf J}_D|$ (lower panel) for stellar mass ratios in the range [1: 100] as indicated by the different colors. For all simulations $\beta_p=0\degr$. The angle $\gamma_p$ of the stellar orbit is 45\degr, 90 \degr, 135\degr and 180\degr. The black horizontal lines indicate the total mass and angular momentum of the initial disc. }
\label{fig:rp_1,i_0,m,l}
\end{figure}

\begin{figure}
\centering
\includegraphics[width=8cm]{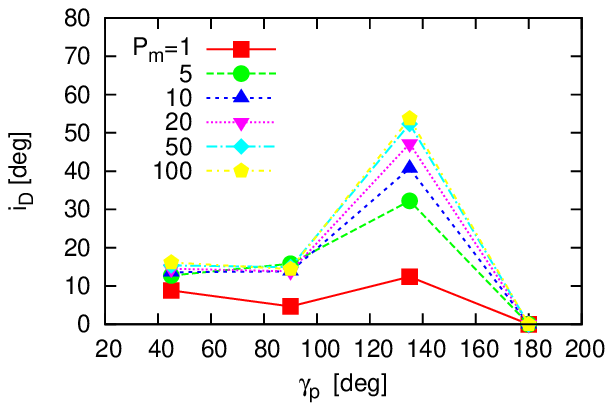} 
\includegraphics[width=8cm]{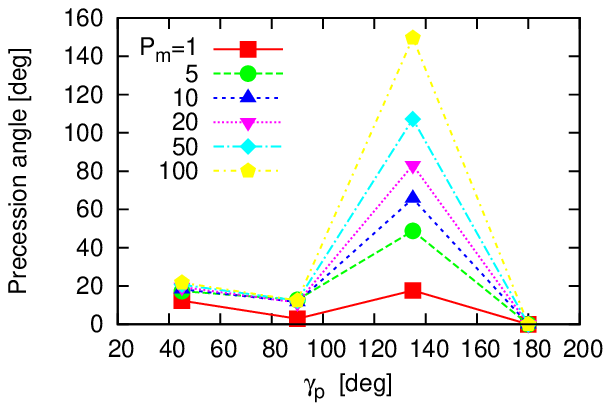} 
\includegraphics[width=8cm]{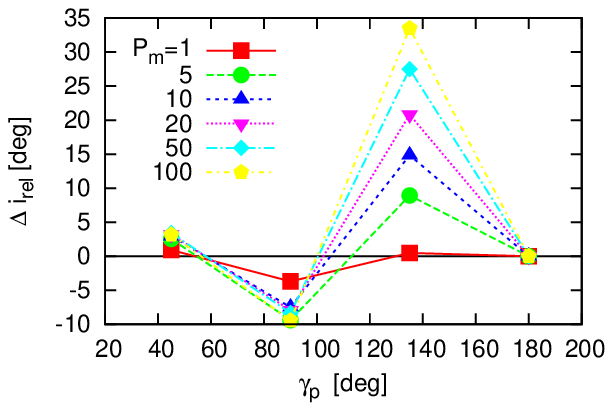}
\caption{ Final disc inclination $i_D$ with respect to the $(x,y)$ plane (upper panel), precession angle (middle panel) and change of relative inclination for stellar mass ratios in the range [1: 100] as indicated by the different colors. For all simulations $\beta_p=0\degr$. The angle $\gamma_p$ of the stellar orbit is 45\degr, 90 \degr, 135\degr and 180\degr.  }
\label{fig:rp_1,i_0,i}
\end{figure}

We now study parabolic orbits which are rotated about the $y$-axis as shown in the lower panel of Fig. \ref{fig:orbit}. In these cases, the pericenter of the perturber is moved out of the initial disc's plane.

Fig. \ref{fig:rp_1,i_0,m,l} shows the final total remaining mass (upper panel) and angular momentum (lower panel) of the disc for simulations with $q=1$ and $\beta_p=0\degr$. Stellar mass ratios in the range [1: 100] are considered as well as different values for $\gamma_p$. 

Prograde orbits with $\gamma_p<90\degr$ show a significantly higher disc mass and angular momentum loss than retrograde cases.  
The results reveal a close similarity to the results of $\gamma_p=0$ and $\beta_p\neq 0\degr$ although the orientations of the stellar orbits are different. 
Prograde orbits with $\gamma_p<90\degr$ show a significantly higher disc mass and angular momentum loss than retrograde cases.  


Figure \ref{fig:rp_1,i_0,i} shows the corresponding final disc inclination $i_D$ with respect to the $(x,y)$ plane (upper panel), the precession angle (middle panel) and the change of the relative inclination (lower panel) for the simulations shown in Figure \ref{fig:rp_1,i_0,m,l}. 
In these cases with $\beta_p=0$ and $\gamma_p \neq 0$, the initial relative inclination between the disc angular momentum vector and the orbital angular momentum vector of the perturber is $i_{rel,0}=\gamma_p$. 

Again, the retrograde case with $\gamma_p=135\degr$ shows the highest final disc inclinations. For $P_m\geq 50$, the inclination is $\sim 50\degr$. $i_D$ decreases with decreasing $P_m$ because of the reduced gravitational influence of the perturber. In the prograde case of $\gamma_p=45\degr$, the perturber is not capable to generate disc inclinations above $\sim 15\degr$.

We again point out the significant disc mass loss in the prograde case which does not allow a large final disc inclination whereas in the retrograde case, the disc mass loss is comparable small. This results in a large final disc inclination.  
The conclusions made in the previous subsection are also valid. 


The study of the precession angle and the change of the relative inclination between the disc and the perturber's orbit reveals the significance of orbital inclination change in the cases $\gamma_p=90\degr$ and $\gamma_p=135\degr$. 
In both cases, the final disc inclination is determined by both the precession and orbital inclination change. In the retrograde case of $\gamma_p=135\degr$, the relative inclination can change by several tens of degrees. This allows the conclusion that precession of ${\bf J}_D$ about ${\bf J}_p$ is not the only dominant process which shapes the final disc inclination.

The general outcome is the same as in the case of ($\beta_p=0$, $\gamma_p \neq 0$). Given the results for the two different rotations, the way of tilting (about $x$- or $y$-axis) the perturber's orbit does not have a significant impact on the final disc properties. It is the magnitude of the  inclination angle $\beta_p$ and $\gamma_p$, respectively, which determines the fate of the gas particles.

\subsection{$q=1$, $P_m=10$, Survey of $\beta_p$ and $\gamma_p$ }

In order to cover the parameter space with both $\beta_p$ and $\gamma_p$ being $\neq 0\degr$, we show a survey with different combinations of ($\beta_p, \gamma_p$).
For this survey, we concentrate on the case $q=1$ and $P_m=10$ only. 
The orbit of the perturber is first rotated about the $x$-axis by $\beta_p$ and then about the $y$-axis by $\gamma_p$.  

\begin{figure}   
\centering
\includegraphics[width=8cm]{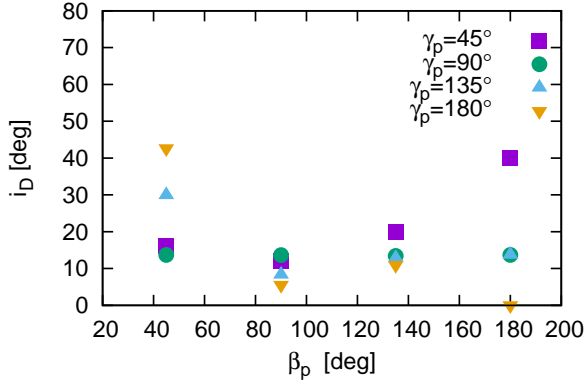} 
\caption{Final disc inclination with respect to the $(x,y)$ plane for a  stellar mass ratio $P_m=10$. The different colored points indicate different values of $\gamma_p$. The values of $\beta_p$ are in the range $[45\degr: 180\degr]$.  }
\label{fig:rp_1,i,gamma,i}
\end{figure}

Figure \ref{fig:rp_1,i,gamma,i} shows the final disc inclination with respect to the $(x,y)$ plane for  $q=1$ and $P_m=10$ and different combinations of $\gamma_p$ and $\beta_p$. 
The case of $\gamma_p=90\degr$ (green filles circules) shows for all values of $\beta_p$ in the range $[45:180]\degr$ exactly the same result. This is because in all four simulations, the perturber's orbit is perpendicular to the initial disc's plane. Due to the axisymmetric nature of the disc, the perturber's orbit relative to the disc is the same in all four cases.

The largest final disc inclinations are found for $(\beta_p,\gamma_p)=(45\degr, 180\degr)$, $(180\degr,45\degr)$ and $(45\degr, 135\degr)$.
For these configurations, the sequence of rotations about the $x$- and $y$-axis results in a retrograde inclined orbit.
All other parameter pairs only allow disc inclinations below $20\degr$. $\beta_p=180\degr$ alone only represents a retrograde coplanar orbit. However, by tilting the orbit about the $y$-axis 45\degr in a second step, the final perturber's orbit is made retrograde and inclined. This is the most favourable configuration for inclination generation. 
In contrast, in the case of $(\beta_p, \gamma_p)=(135, 135)\degr$, an initial retrograde inclined orbit with $\beta_p=135\degr$ is rotated about the $y$-axis by $135\degr$ which results in a final prograde orbit. 

Taken all results together, the clear conclusion for single stellar flybys is that the perturber has to be on a retrograde inclined orbit relative to the initial disc orientation. As long as the motion of the perturber is retrograde and inclined, it is able to generate significantly high inclinations in the gaseous disc, independently on its specific orientation to the initial disc.

\subsection{Global structure of disc} \label{sec:longterm}

\begin{figure} 
\centering
\includegraphics[width=7.5cm]{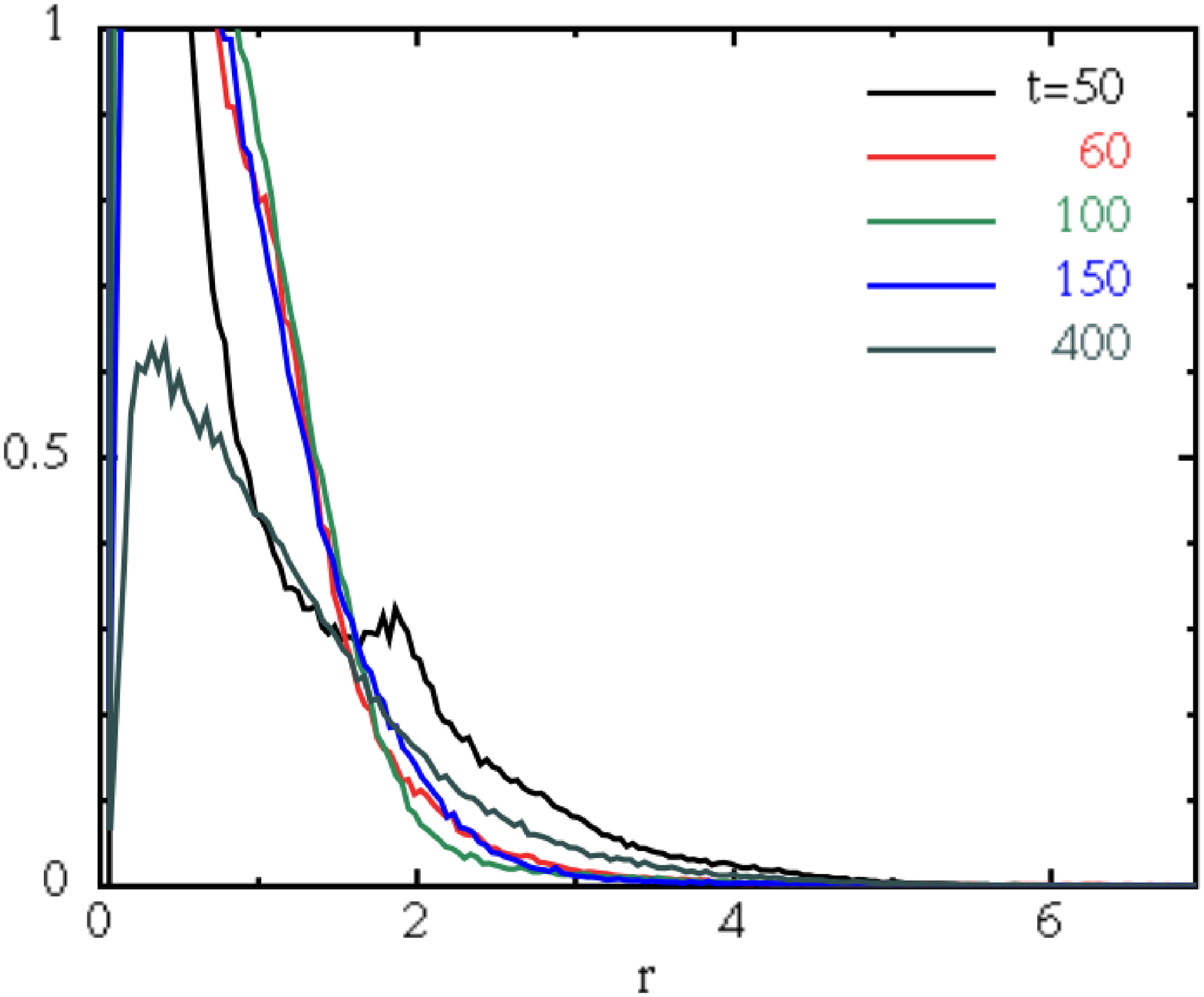} \\
\includegraphics[width=8cm]{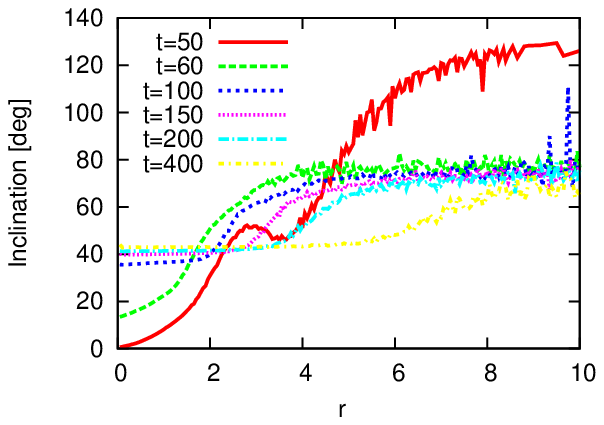}
\caption{Upper panel: The azimuthally averaged disc surface mass density profile (in units of $M_\rmn{J}/(5\rmn{au})^2$); lower panel: averaged inclination $i(r)$ of gas particles as a function of distance (in units of 5 au) for different time steps and $P_m=10$, $q=1$, $\beta_p=135\degr$, $\gamma=0\degr$. [The surface density profile is created using SPLASH \citep[][]{Pri2007}.]}
\label{fig:structure}
\end{figure}

\begin{figure} 
\centering
\includegraphics[width=7.5cm]{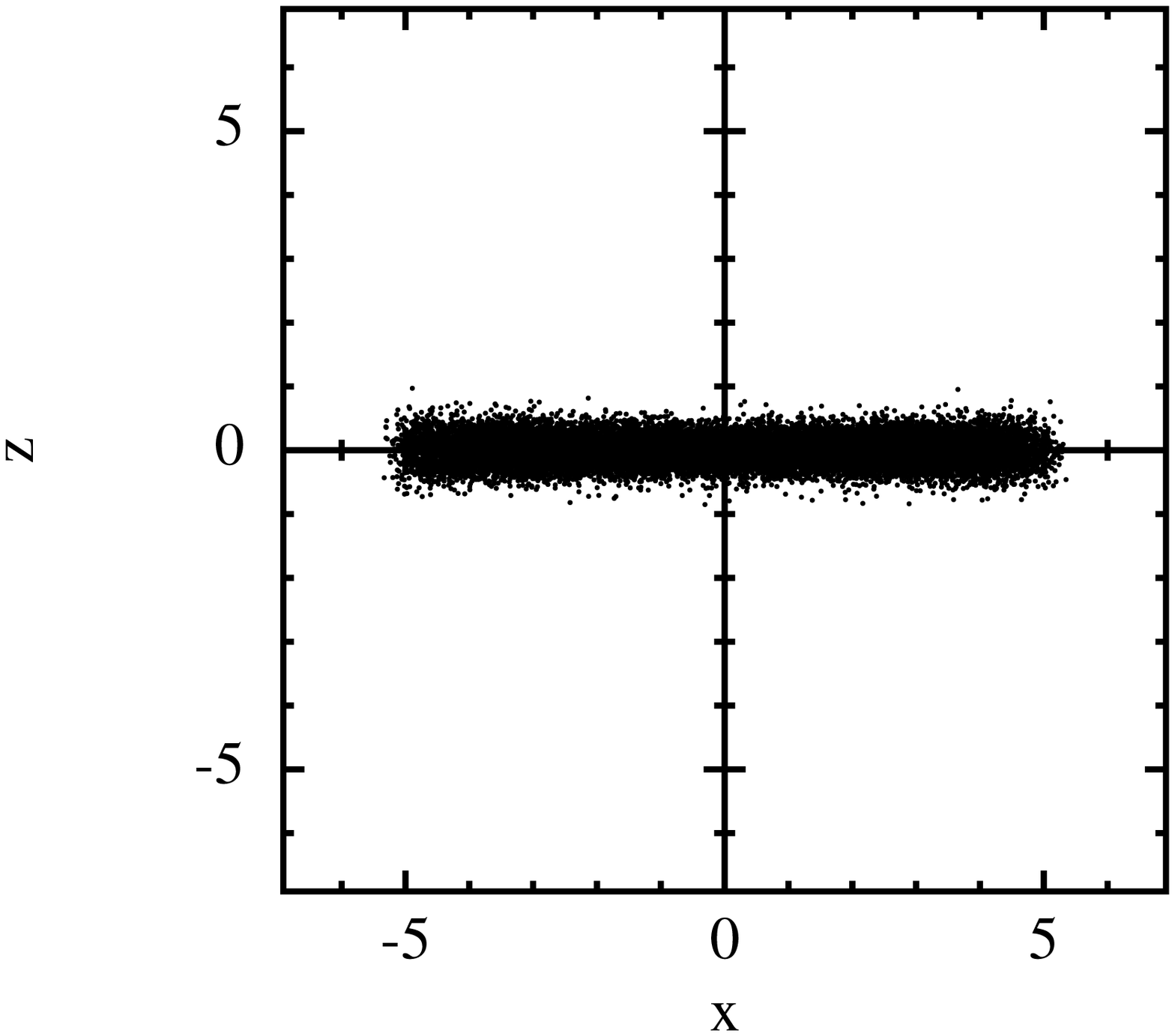}
\includegraphics[width=7.5cm]{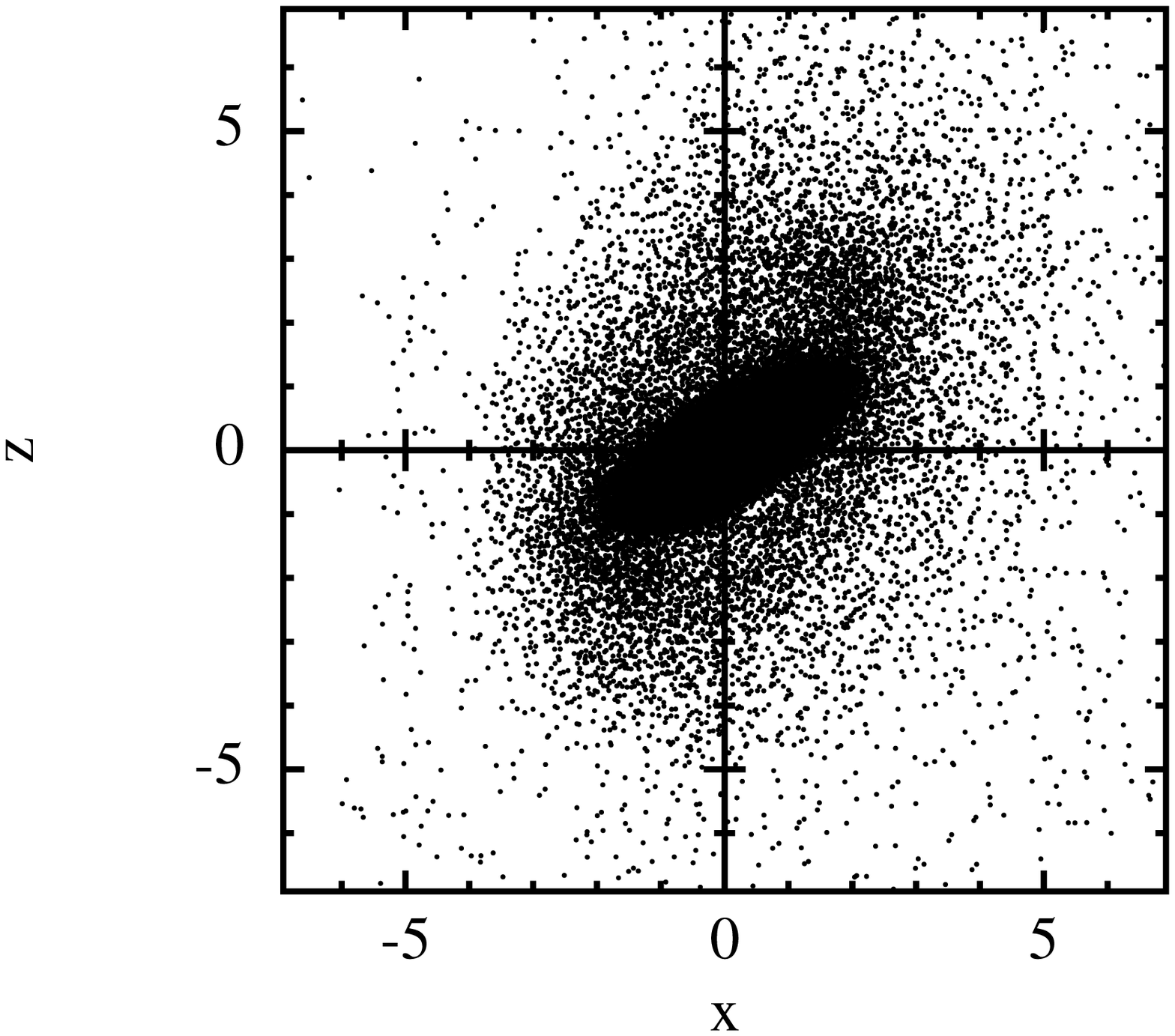}  
\caption{$xz$- view of the particle distribution at time $t=0$ (upper panel) and $t=100$ (lower panel) for the same simulation as shown in Fig. \ref{fig:structure}. }
\label{fig:xz}
\end{figure}

\begin{figure} 
\centering
\includegraphics[width=7.5cm]{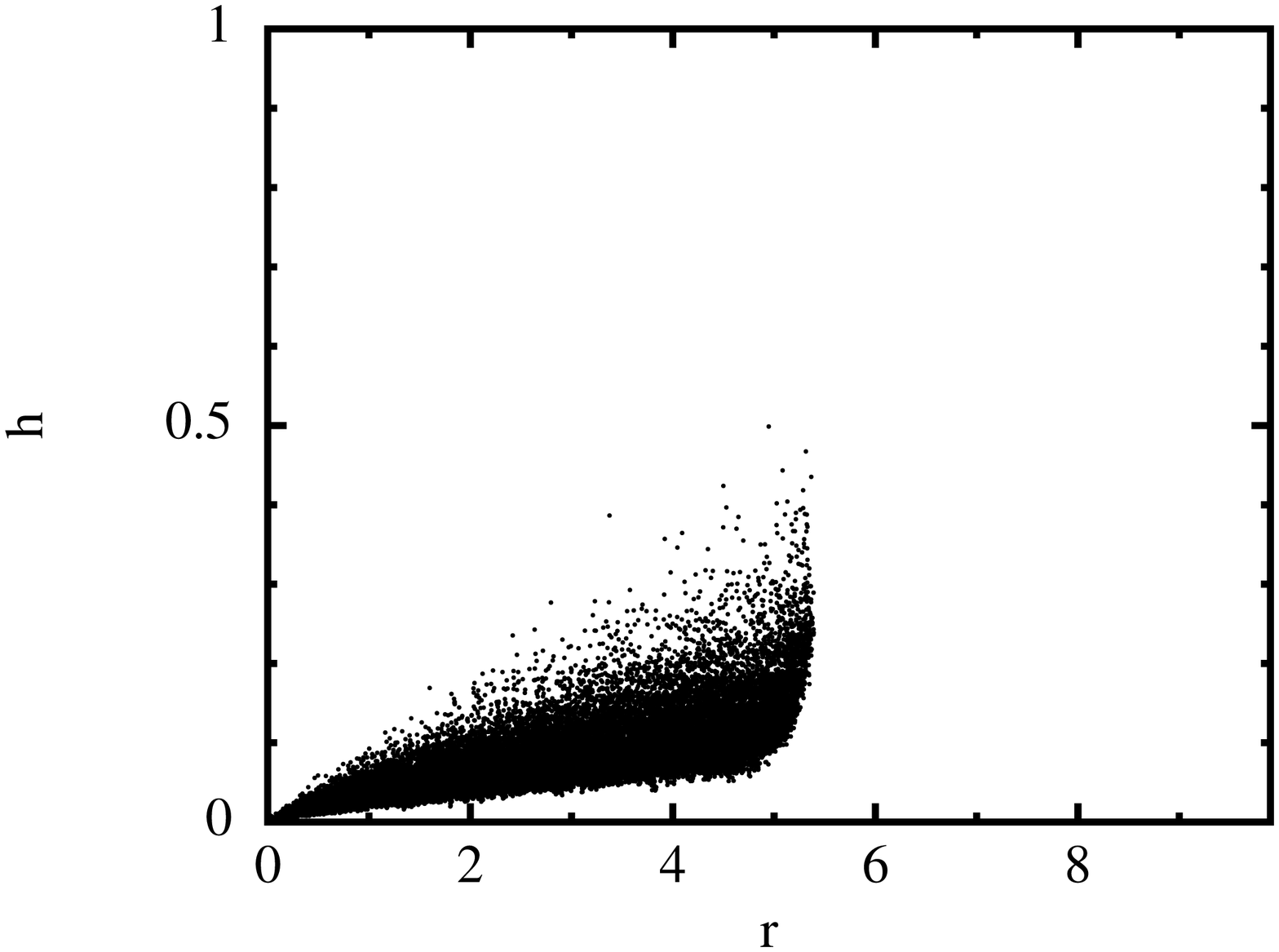} 
\includegraphics[width=7.5cm]{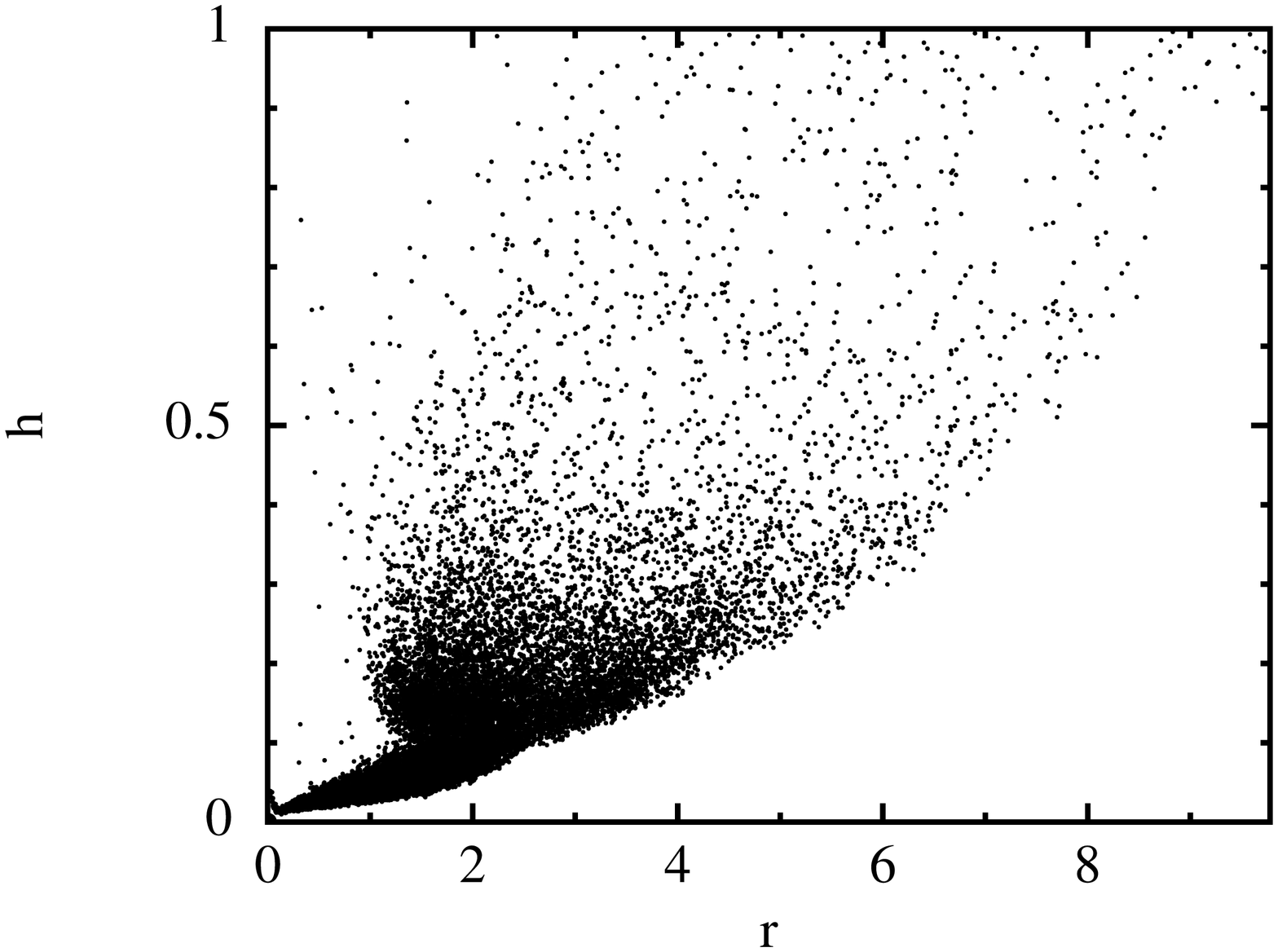} 
\includegraphics[width=7.5cm]{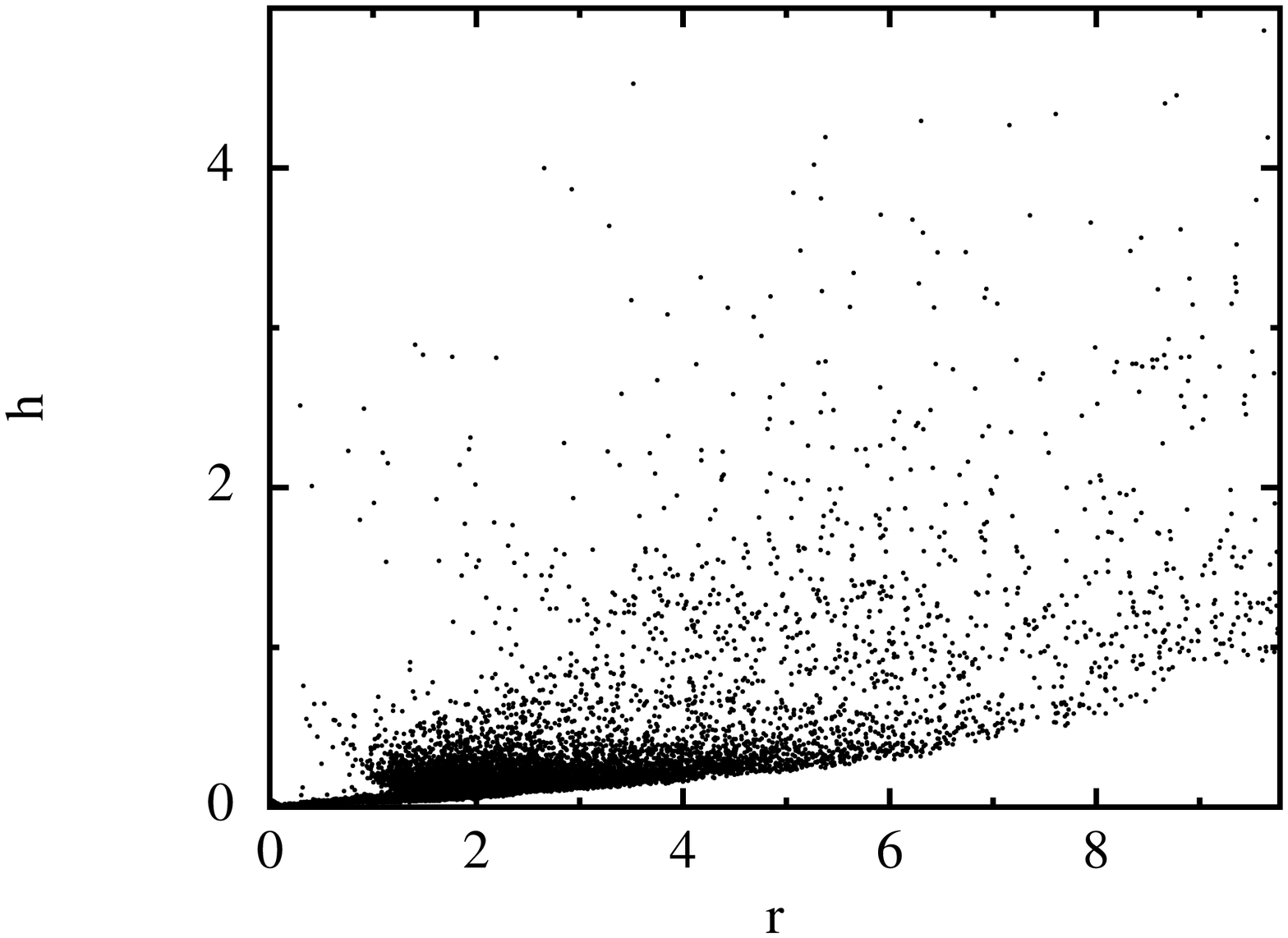} 
\caption{Smoothing length as function of distance to the center at time $t=0$ (upper panel) and $t=100$ (middle and lower panel) for the same simulation as shown in Fig.
\ref{fig:structure}. Note the increased $y$-range in the lower panel. }
\label{fig:h}
\end{figure}

\begin{figure} 
\centering
\includegraphics[width=7.5cm]{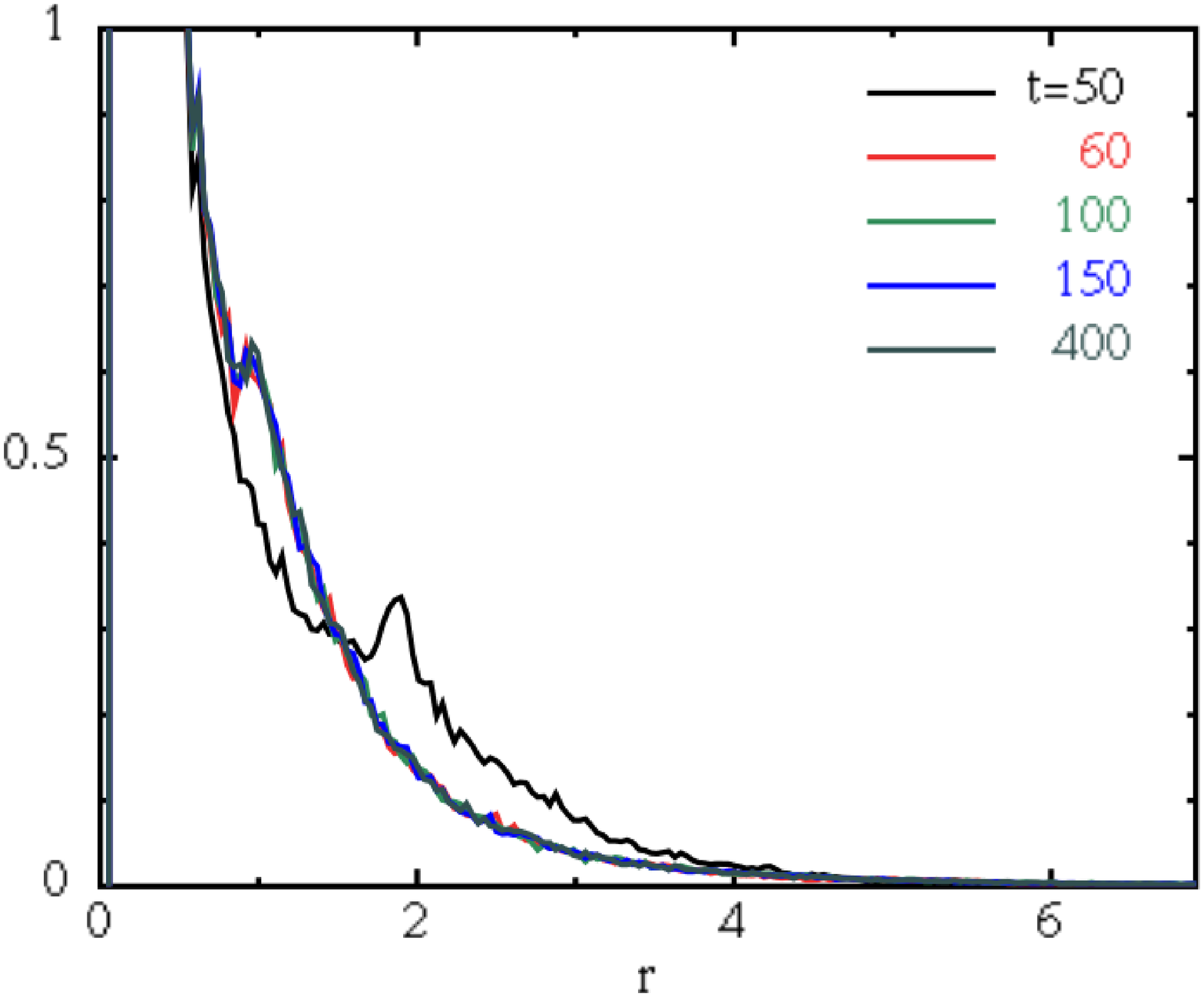}\\ 
\includegraphics[width=8cm]{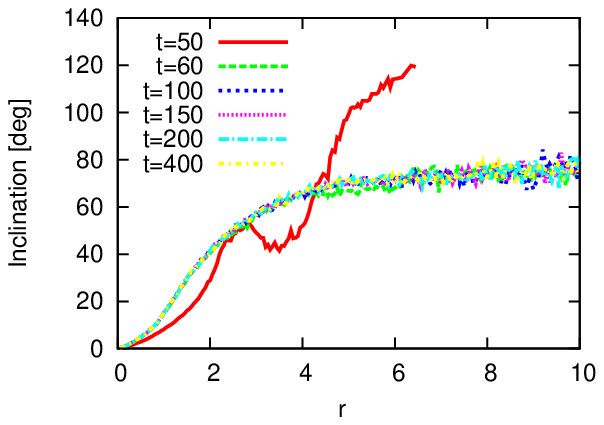}
\caption{The azimuthally averaged disc surface mass density profile (in units of $M_\rmn{J}/(5\rmn{au})^2$) and inclination of gas particles as a function of distance (in units of 5 au) for different time steps for a purely N-body simulation of the disc. The simulation parameters are   $P_m=10$, $q=1$, $\beta_p=135\degr$, $\gamma=0\degr$. [The surface density profile is created using SPLASH \citep[][]{Pri2007}.]}
\label{fig:structure_Nbody}
\end{figure}

 Figure \ref{fig:structure} shows the azimuthally averaged surface mass density of the disc (upper panel) and the averaged inclination of gas particles (lower panel) for timesteps in the range $t=50 - 400$ orbits and distances up to $r=10$ internal length units. 

The averaged inclination $i(r)$ of gas particles is determined by averaging over all particles inside a shell with a radial width of $\Delta r=0.05$. The parameters for this simulation are $P_m=10$, $q=1$, $\beta_p=135\degr$ and $\gamma=0\degr$.

The initial disc size is $r=5$. For $r>5$, the surface mass density $\Sigma \to 0$. The inclination at large $r$ can be high as can be seen in the lower plot. We have also studied the inclination beyond $r=10$, but found no significant variations of $i(r)$ there.  

The first timestep shown is $t=50$ which is the time of pericenter passage. At this timestep, the disc surface mass density differs  significantly from its initial profile and it shows a local maximum at $r=2$. The inclination plot for $t=50$ shows that the disc has generated significant inclinations everywhere with the highest inclinations up to $\sim$ $130\degr$ in the outer parts with $r \to 10$. 

In the following timesteps, the absolute magnitude of the local maximum in the surface mass density increases and moves inwards. This is a result of angular momentum loss of the gas particles to the perturber which results in an inward motion of the gas particles. At $t>150$, most of the disc mass is inside $r=4$.

The evolution between $t=50$ and $t=60$ is characterized by an inclination growth in the inner part and inclination decrease in the outer part. The evolution during this relatively short time period is mainly determined by angular momentum loss of gas particles. As we have shown in Fig. \ref{fig:rp_1,gamma_0,example_1}, the disc loses a significant fraction of its initial angular momentum. As a result, the gas particles rearrange and the highly inclined particles in the outer parts move inwards. This inward motion of the gas particles is significant between $t=50$ and $t=60$ as the surface mass density profiles of these two timesteps differ strongly. The inward motion of highly inclined gas particles results in a growth of inclination in the inner regions, while in the outer region, the disc inclination decreases because of lack of highly inclined particles. In addition, there are also particles with high orbital inclinations in the outer regions which become bound to the perturber and thus leave the system. 

With respect to the long-term evolution of the disc inclination distribution, an additional mechanism dominates the behaviour of the disc. The second mechanism which operates in the gaseous medium is the mutual interaction between the gas particles  through hydrodynamical forces. The hydrodynamical treatment of the disc results in angular momentum transfer between the gas particles and large regions of constant orbital inclination inside the disc which will be described in detail in the following. These two mechanisms taken together can fully explain the evolution of the disc inclination.

After the close encounter with the passing star, the disc quickly (in less than $\Delta t \leq 20$ orbits) settles into a quasi-steady state which can be described by a warped structure. In this warped structure, two radial domains can be identified with significantly different disc inclinations. Note that inside both radial domains, the disc inclination is constant because of the communication of gas particles inside the disc.
 In the example shown in Fig. \ref{fig:structure}, the inner part of the disc is less inclined than the outer part because of the stronger influence of the perturbing star on the outer disc particles. 

For $t>60$, both the inclination of the inner and the outer part do not change significantly anymore. 
However, the hydrodynamical evolution of the disc increases the inner region with constant inclination. In addition, the initial sharp transition between the two radial domains is smeared out so that at $t=400$, there is a slow and smooth transition between the two domains instead of the sharp edge at smaller times.

 As mentioned in Section \ref{sec:artv}, the Shakura Sunyaev parameter for our code has been tested in the so-called Ring spreading test \citep[see][for more details]{Xia2013} to be roughly 0.02. However, this value only applies to normal Keplerian discs without significant extent in $z$-direction and cannot explain the observed viscous evolution of the disc in Fig. \ref{fig:structure} which takes place on a very short timescale of a few 100 orbits.  In the scenario of a single stellar flyby, many SPH particles (mostly in the outer disc regions) are scattered in the 3D space into individual inclined orbits as can be shown in Fig. \ref{fig:xz} for $t=60$. 
In this case, the numerical resolution decreases very rapidly. In regions with $r>2$, the small local densities produce very large smoothing lengths (Fig. \ref{fig:h}) which are the key parameters for the computation of all hydrodynamical quantities such as pressure and artificial viscosity. Compared to the initial smoothing lengths (upper panel of Fig. \ref{fig:h}), the smoothing lengths at later times reveal a much broader spreading into very large values. 
The lower panel of Fig. \ref{fig:h} shows that smoothing lengths of  particles can even reach several internal length units which is comparable to the extent of the initial disc with a radius of 5 internal length units. 

As a consequence, the artificial viscosity in these low resolutional regions is unphysically large. For example, in the case of a 100 AU disc, a smoothing length $h$ of 1 length unit would imply a sphere with radius 20 AU within which only 40 SPH particles are present. Thus, the hydrodynamical quantities for such a SPH particle with $h=1$ are computed by considering an unphysically large region with a diameter of 40 AU.

In these extremely low resolution regions, the Shakura Sunyaev parameter is not 0.02 anymore. The viscosity is extremely overestimated making the viscous evolution in the outer disc regions much faster. For this reason, both the inward motion of the particles as well as the smearing out of the viscous disc happen on a short timescale which is mainly due to the low numerical resolution in the outer regions. 

The increase of SPH particles cannot solve this problem, but rather shifts it to larger radii.
A solution for this numerical problem could be the implementation of test loops in the code to check the adequacy of the smoothing length of a SPH particle. In the case of insufficient resolution, the particle could be treated as a purely N-body particle by ignoring the hydrodynamical terms. However, the reinclusion of the particle into the hydrodynamical fluid has to be considered, too. These modifications should be investigated with more detail.   

Given the current simulation results, we point out that the outer regions of an initial thin disc could be strongly affected by a stellar perturber making an appropriate hydrodynamical description of the gas particles challenging. It has to be investigated in a subsequent work to which extent the gas particles can still be studied as a hydrodynamical entity.   


To summarize, directly after the stellar flyby, the disc evolves into a warped structure with the inner part of the disc being less inclined than the outer part of the disc. Both the inner and the outer part have constant inclinations while the intermediate region shows a sharp transition between the two inclined regions.  This warped structure is smeared out by the viscous disc so that at the end of our simulation, the disc inclination is dominated by the inner part.
But note that for large radii, the numerical resolution of the SPH method is insufficient due to small mass densities. Thus, the accurate time-dependent evolution of the warped structure cannot be described appropriately by our simulations.

Most of the disc mass is concentrated in the inner region. At $t=200$, more than 70\% of the total disc mass is located inside $r=2$. Because of gas accretion in the inner region, the mass fraction inside $r=2$ decreases to 50\% at $t=400$. 
Thus, the relevant part of the disc which defines the total inclination and dynamics of the disc is located in the inner region.

For comparison, we show the azimuthally averaged surface mass density of the disc (upper panel) and the averaged inclination of gas particles (lower panel) for a purely N-body simulation and timesteps in the range $t=50 - 400$ internal time units in Fig. \ref{fig:structure_Nbody}. The parameters are chosen to be the same as in the hydrodynamical simulation shown in Fig. \ref{fig:structure}.

Again, we show the radial domain up to $r=10$ within which all disc particles are located.
The only difference between the N-body and the SPH simulation is the lack of hydrodynamical evolution and thus no communication between the gas particles in the N-body case.

At $t=50$, the surface mass density profile of the N-body run is similar to the SPH case. The density in the vicinity of the pericenter of the perturber is slightly higher than in the SPH case because in the SPH simulation, the particles are bound to the each other and thus cannot move as freely as in the N-body case. 

Due to angular momentum loss to the perturber, the gas particles move inwards and settle down on their new orbits within 10 time units as can be seen in the new density profile with a higher mass amount in the inner region. During this period, we also observe an inclination increase in the inner regions and an inclination decrease in the outer regions which, as discussed in the SPH case, is a result of angular momentum loss of the disc particles and removal of particles by the perturber. 
As there are no additional angular momentum changes of the particles, the final disc configuration is reached at $t=60$. In this context, we note that \citet{Bha2015} performed a more complete N-body parameter study of inclined parabolic perturber orbits and their influences on massless test particles in order to study the disc sizes of protoplanetary discs.

The striking difference between the N-body and SPH simulation is the absence of a warped structure in the N-body simulation. The warping can only be produced if gas particles are allowed to communicate with each other.
Hence, the inclusion of hydrodynamical interaction of gas particles inside the disc is crucial when the evolution of inclination inside a protoplanetary disc is studied. Dependent on the evolutionary stage of a disc which determines whether hydrodynamical interaction is still active or not, different outcomes can be gained which lead to different radial inclination distributions inside the disc.  

These results have direct consequences on the formation of misaligned planets. As we have shown with the numerical simulations, a single stellar flyby which generates inclination in the outer parts of a viscous disc can trigger the formation of a warped disc and a significantly inclined inner disc. For such a warped disc, the inclination of the inner part can be as high as several tens of degrees. 

Planets that are formed in the inner region are naturally expected to have an initial orbital inclination as high as the inner disc inclination. 
The scenario of a stellar flyby can therefore be understood as a method to produce inclined protoplanetary discs which, at a later stage, form planets with misalignment angles of up to $\sim 60\degr$. Subsequent planet migration could lead to the formation of misaligned Hot Jupiters.

The angle with the highest final disc inclination is determined to be $\gamma_p=135\degr$ which is a retrograde orbit with an inclination of 45\degr. Note that the angle with the maximum final disc inclination is expected to depend on the specific parameters and might vary for different parameter sets.

\section{Conclusions} \label{sec:conclusions}

The aim of this work was to test whether single stellar flybys are capable to generate significant disc inclinations which would allow the natural formation of misaligned planets and to identify the parameter sets associated with high final disc inclinations.
For this, we have performed  SPH simulations in order to study the influence of single stellar flybys on the evolution of viscous protoplanetary discs. 
We have run simulations with a stellar pertuber being on a parabolic inclined orbit and the stellar mass ratio $P_m=M_p/M_*$ in the range $[1:100]$, the impact parameter (periastron distance relative to the outer disc radius) $q=r_p/R_{\rmn{out}}=[1:10]$ and the two angles $\beta_p$ and $\gamma_p$ which characterize the three-dimensional orientation for the perturber's orbit. These parameter ranges enabled us to cover a large parameter space for stellar flybys. 

We first studied the two extreme cases of $(\beta_p=0\degr$, $\gamma_p\neq 0\degr)$ and $(\beta_p\neq 0\degr, \gamma_p=0\degr)$. Both cases showed the same outcome with respect to the disc evolution. A perturber on a prograde inclined orbit is able to remove a large fraction of the initial disc mass and angular momentum. This result agrees well with previous works done by e.g. \citet{Ost1994, Pfa2005, Olc2006} which showed that prograde orbits are most destructive.
This high disc mass loss has direct impact on the total final disc inclinations. In the prograde cases, a stellar flyby fails to generate any significant disc inclinations independently on the perturber's mass since all particles with high orbital inclinations are removed from the initial disc. Those particles which have not been significantly disturbed by the perturber only harbour minor orbital inclinations. 

For retrograde orbits, the disc mass loss and angular momentum loss are much less than in the prograde case. In these cases, significant disc inclinations up to 60\degr could be generated. Thereby, the final disc inclination decreases with increasing impact parameter $q$ and decreasing stellar mass ratio $P_m$.  

In contrast to previous studies of stellar flybys with sufficiently small perturbations to which a disc is able to response linearly, our simulations have to be assigned to the non-linear regime. In this regime, the final orientation of a gas disc is not only determined by precession of its angular momentum ${\bf J}_D$ about the orbital angular momentum vector of the perturber ${\bf J}_p$ as is the case in the linear regime. But it is rather determined by both the precession and the orbital inclination change of the disc. In the case of rigid body precession without change of orbital inclination, it is expected that the relative inclination $i_{rel}$ between ${\bf J}_D$ and ${\bf J}_p$ should stay constant. However, in the most extreme cases of the perturber being on a retrograde inclined orbit, our simulations have revealed relative inclination changes by several tens of degrees.  
This is a clear indication that during stellar flybys with strong perturbations, orbital changes of the disc make a significant contribution to the final disc orientation.

In a next study, we have performed simulations with both $\beta_p$ and $\gamma_p$ being $\neq 0\degr$. The results suggest that for given stellar mass ratio and impact factor, the most important condition for the generation of high disc inclinations is that the perturber's orbit is retrograde and inclined. The specific orientation of the pertuber's orbit thereby does not play a major role. The condition of a retrograde and inclined orbit and the arbitrariness of the specific orientation taken together make it very probable for close encounters to generate disc inclinations. But note that because of significant mass removal, disc inclinations above $60\degr$ could not be generated by a single stellar flyby. 

A detailed study of the internal form of the viscous disc has revealed a clear warping. Two different radial domains occured with significantly different orbital inclinations. While in the outer parts, the disc inclination was very high and the disc mass was very low, most of the disc mass was concentrated in the inner region with $r<4$ where disc inclinations could still reach values above 50\degr. We also remark that due to insufficient resolution in the outer parts, the hydrodynamical description is not appropriate and alternative methods should be introduced in future work.  More precisely, with the numerical work shown in this paper, we succeeded in simulating the formation of inclined and warped discs. However, their long-term evolution cannot be described accurately due to insufficient numerical resolution of widely spreaded particles in the three-dimensional space. The insufficient resolution resulted in an extreme overestimation of the artificial viscosity which led to unphysically fast viscous 
evolution of the gas medium.


In connection with misaligned Hot Jupiters, our results suggest that single flybys can be responsible for the existence of Hot Jupiters with misalignment angles up to 60\degr. For these angles, stellar flybys with realistic mass ratios of $P_m \gtrsim 5$ and impact parameters as small as $q\leq 3$ are required.

In Section \ref{sec:mass}, we have determined the probability of a solar-like star to have at least one close encounter with a massive star on a retrograde inclined orbit to be $\sim$ 0.125. 
Given this relatively high probability of 12.5 per cent for close retrograde encounters, large disc inclinations or misaligned planetary orbits could be a result of stellar flybys. 
Since single stellar flybys can result in orbital inclinations in the range [0\degr:60\degr], a large fraction of observed misaligned Hot Jupiters with angles below $60\degr$ could be the result of only one single stellar flyby. Although the scenario of single flybys fails to explain the formation of planets with even higher misalignment angles, it could be possible that a sequence of stellar flybys is responsible for their existence. The scenario of several stellar flybys has to be studied in a future project. 


Another mechanism which involves the generation of high disc inclinations with the help of stellar flybys is the inclusion of possible mass accretion between the gas discs around two  stars.
In the simulations shown in this paper, we have only studied one gaseous disc around the primary star. In stellar clusters, all stars are generally expected to possess an own gaseous envelope/disc. The gas disc around the passing star has not been included in our simulations. For close encounters, mass accretion from one disc to the other can play a major role in understanding the evolution of the disc inclination \citep[e.g.][]{Thi2011}. However, in order to understand subsequent results of stellar flybys involving two protoplanetary discs and make comparisons between the results, it is of great importance to firstly understand the effects of a single stellar flyby without mass accretion for which we have made major contributions in this paper. Another justification for our survey without mass accretion is the possibility that a star has lost its gaseous envelope during an earlier close encounter so that for  future encounters, it only appears as one single massive body. 
By including a second protoplanetary disc, additional parameters have to be included into the study. As the orientation of the second disc is not necessarily correlated to the orbit of the second star, different orientations of the second disc have to be studied for a specific orbit of the perturber. Close encounters between stars with different stellar mass ratios $P_m$ and two gas discs with arbitrary orientations will be studied in a subsequent paper.

To summarize, single stellar flybys with the perturber being on a parabolic retrograde inclined orbit are capable to generate significant disc inclinations and hence provide the natural formation of misaligned Hot Jupiters. In consideration of the numerical survey, this scenario is very promising in explaining misalignment angles up to 60\degr. However, our proposed scenario fails to generate higher misalignment angles with $i_D>60\degr$ because of the large amount of disc mass loss to the perturber.
For highly inclined or even retrograde Hot Jupiters, alternative formation scenarios have to be studied such as mass accretion or a sequence of stellar flybys. These scenarios could help to fully understand the formation of misaligned Hot Jupiters.


\section*{Acknowledgments} 
The author gratefully acknowledges the helpful discussions with Susanne Pfalzner.
Xiang-Gruess acknowledges support of the Max Planck Society through a postdoctoral fellowship.
We acknowledge the computing time granted (NIC project number 8163) on the supercomputer JUROPA at J\"ulich Supercomputing Centre (JSC) .



  



\label{lastpage}

\end{document}